\documentclass[12pt,twoside]{article}
\pdfoutput=1

\newlength{\dinwidth}
\newlength{\dinmargin}
\usepackage{amsmath, amsthm, amsfonts, amssymb, bm, mathrsfs, indentfirst,
sectsty, graphicx, fancyhdr, slashed, fullpage, color, authblk}

\begin{document}

\begin{titlepage}

\flushright{IFT-UAM/CSIC-13-086}\\[2cm]
\begin{center}
 {\Large \bf \sc Holographic Superfluids and the Landau Criterion}
\\[1.52cm]
{\large Irene Amado$^{a,}$\footnote{irene.r.amado@gmail.com}, Daniel
Are\'an$^{b,e,}$\footnote{arean@sissa.it},
Amadeo Jim\'enez-Alba$^{c,}$\footnote{amadeo.j@gmail.com},\\
Karl Landsteiner$^{c,}$\footnote{karl.landsteiner@csic.es}, Luis
Melgar$^{c,}$\footnote{luis.melgar@csic.es} and Ignacio Salazar
Landea$^{d,b,}$\footnote{peznacho@gmail.com} }

\bigskip

{}$^{a}${\small \it Department of Physics, Technion, Israel Institute of
Technology, Haifa 32000, Israel}\\
{}$^{b}${\small \it International Centre for Theoretical Physics
(ICTP), Strada Costiera 11; I 34014 Trieste, Italy}\\
{}$^{c}${\small \it Instituto de F\'\i sica Te\'orica IFT-UAM/CSIC, C/ Nicol\'as
Cabrera 13-15, Universidad
Aut\'onoma de Madrid, 
28049 Cantoblanco, Spain}\\
{}$^{d}${\small \it Instituto de F\'\i sica La Plata (IFLP) and
Departamento de F\'\i sica Universidad Nacional de La Plata, CC 67,
1900 La Plata, Argentina}\\
 {}$^{e}${\small \it INFN - Sezione di
Trieste, Strada Costiera 11; I 34014 Trieste, Italy}
\end{center}

\bigskip

\begin{abstract}
We revisit the question of stability of holographic superfluids with finite
superfluid velocity. Our method is based on applying the Landau criterion to the
Quasinormal Mode (QNM) spectrum. In particular we study the QNMs related to the
Goldstone modes of spontaneous symmetry breaking with linear and quadratic
dispersions.
In the linear case we show that the sound velocity becomes
negative for large enough superfluid velocity and that the imaginary part of the
quasinormal frequency moves to the upper half plane. Since the instability 
is strongest at finite wavelength, we take this as an indication 
for the existence of an inhomogeneous or striped condensed phase for large
superfluid velocity. In the quadratic case the instability is present for
arbitrarily small superfluid velocity. 
\end{abstract}
\end{titlepage}


\section{Introduction}
\setcounter{footnote}{0}

The characteristic property of a superfluid is its ability to flow totally
frictionless through thin capillaries. It is useful to think of a superfluid
as a two component liquid. One component is the ground state with a macroscopic
occupation number and the other is the normal component, subject to friction
and viscosity. At very low temperatures the normal component can be described as
the gas of elementary quasi particle excitations above the macroscopically
occupied ground state. A famous argument due to Landau
\cite{Landau,Khalatnikov,PinesNozieres} sets a limit to the flow velocity that
the condensate can obtain. The essence of the argument is as follows. At zero
temperature the energy of a quasiparticle excitation of momentum $\vec{p}$ is
$\epsilon(\vec p)$ in the rest frame of the condensate. If we imagine a
situation in which the condensate moves with constant velocity $\vec{v}$ the
energy cost in creating a quasiparticle is 
\begin{equation}\label{eq:energyboost}
 \epsilon'(\vec p ) = \epsilon(\vec{p}) + \vec{v} \cdot \vec{p}\,.
\end{equation}
In particular if $\vec p$ is anti-parallel to the flow velocity $\vec v$ this
energy is diminished and eventually goes to zero. If $\epsilon' <0$ it is
energetically favorable for the system to create elementary excitations and
populate states with this effective negative energy. Since the superfluid
velocity $\vec v$ is kept constant this means that eventually the condensate
gets completely depleted and the superflow stops. It follows that there is a
critical flow velocity above which the superfluid ceases to exist. The famous
Landau criterion for the existence of superflow is therefore
\begin{equation}\label{eq:landaucrit}
 v_\mathrm{max} = \min \frac{\epsilon(p)}{p}\,,
\end{equation}
where the minimum over all elementary excitation branches has to be taken. It is
known for example for superfluid helium that the low temperature normal
component can be well described as a gas of phonons and rotons and that the
critical velocity is not determined by the minimum of the phonon and roton
dispersion relation but rather by the excitation of vortices, resulting
in a much lower critical velocity.

At higher temperatures there is always a normal component present and therefore
the energy of an excitation of a superfluid with superflow can not be obtained
by a (Galilean) boost as in equation (\ref{eq:energyboost}). It is however still
true that the energy will depend on the superfluid velocity and that it can
become negative if the superfluid velocity is too large. At finite temperature the
criterion is therefore that the superflow is stable as long as the energy of all
quasiparticle excitations is positive. If in a superfluid the only low energy
excitations are the phonons that criterion is  basically the statement that the
superflow dependent sound velocity is positive for all directions. 

The AdS/CFT correspondence has proven to be a very useful tool for studying
quantum field theories at  strong  coupling. In particular, since we can study
Bose and Fermi systems at finite temperature and chemical potential using
holography, there are many condensed matter physics applications of the duality
(for a review, see \cite{Hartnoll:2009sz,Herzog:2009xv,McGreevy:2009xe}). 

One of the most important achievements of AdS/CMT in the last years is the
construction of geometries dual to superfluids
\cite{Gubser:2008px,Hartnoll:2008vx,Hartnoll:2008kx}. 
The order parameter can be either a scalar, a vector or a spin-2 tensor (we talk
of s-, p-\cite{Gubser:2008wv,Ammon:2008fc} and d-wave superfluidity
\cite{Benini:2010pr,Chen:2010mk}, respectively).

In \cite{Herzog:2008he,Basu:2008st} an s-wave superfluid in 2+1 dimensions with
superflow was constructed  and it was pointed out that there is indeed a
critical velocity above which the superfluid state ceases to exist. The phase
diagram obtained in these works was based on comparing the free energy of the
superflow with the free energy of the normal phase. It turned out that the phase
transition from the superfluid phase to the normal phase was either first or
second order depending on the temperature. Remarkably enough, in 3+1 dimensions
there is some range of masses of the condensate for which the phase transition
is always of second order type \cite{Arean:2010zw}. Another way of establishing
the phase diagram has been used in \cite{Arean:2010xd}. There the supercurrent
was fixed and it was argued that the phase transition is always first order.

The physical significance of the comparison of the free energies of the state
with superflow and the normal state is not totally clear, since for all
temperatures below the critical temperature the normal state is unstable towards
condensation to the superfluid state without superflow. Indeed the superflow by
itself is a metastable state only \cite{PinesNozieres} as emphasized
already in \cite{Herzog:2008he}. We will propose a different method of
characterizing the phase diagram more directly related to the stability
criterion (\ref{eq:landaucrit}).

The purpose of this paper is thus to revisit the question of the realization of
the stability criterion (\ref{eq:landaucrit}) in holographic superfluids. The
simplest holographic models of superfluids are obtained in the so called
decoupling limit. In this limit one discards the fluctuations of the metric and
keeps only the dynamics of a charged scalar field and a gauge field in an
asymptotically AdS black hole. The excitation spectrum of a holographic field
theory at finite temperature and density is given by the spectrum of quasinormal
modes (QNMs) \cite{Horowitz:1999jd, Birmingham:2001pj, Berti:2009kk,
Landsteiner:2012gn}. The QNMs of the simplest holographic superfluid with a
spontaneously broken $U(1)$ symmetry have been obtained in \cite{Amado:2009ts}.
Recently this model has been generalized to a case with $U(2)$ symmetry
\cite{Amado:2013xya}\footnote{This holographic model has also been introduced 
in \cite{Krikun:2012yj}.}, giving rise to the holographic dual of a multi-component
fluid \cite{Halperin}. The spectrum of the $U(2)$ model turned out to contain a
copy of the usual QNM spectrum of the $U(1)$ model but also a novel feature, the
appearance of a type II Goldstone boson.  

It is of course a standard fare that the breaking of a continuous symmetry leads
to the appearance of ungapped states, the Goldstone bosons. This is also
respected by holographic field theories. The Goldstone bosons appear as special
ungapped QNMs. It is less well-known that Goldstone bosons do not necessarily
have linear dispersion relation even in relativistic field theories. Depending
on whether their dispersion relation is proportional to an odd or even power of
the momentum they are called of type I or of type II (see \cite{Brauner:2010wm}
for a review). The appearance of type II Goldstone bosons is also
related to another fact, namely that the number of Goldstone bosons does not
equal the number of broken generators \cite{Miransky:2001tw, Schafer:2001bq,
Watanabe:2013iia}\footnote{Further
recent results on type II Goldstone bosons can be found in
\cite{Kapustin:2012cr, Watanabe:2013uya,Nicolis:2012vf, Nicolis:2013sga}. 
In a holographic context type II Goldstone bosons have been also
found previously in \cite{Filev:2009xp}.}. 
In fact in the holographic model the $U(2)$
symmetry gets broken to $U(1)$ and consequently there are three broken symmetry
generators but only two holographic Goldstone bosons were found. One of them could be
identified with the usual sound mode with linear dispersion relation
\begin{equation}\label{eq:sound}
 \omega(k) = v_s k +(b - i\Gamma)k^2\,,
\end{equation}
where $v_s$ is the speed of sound, $b$ a correction quadratic in momentum and
$\Gamma$ the sound attenuation constant. The type II Goldstone boson on the
other hand was found to have dispersion relation
\begin{equation}
\label{eq:reltypeII} \omega(k) = (B-iC) k^2\,,
\end{equation}
with no linear part. All the constants appearing in these dispersion relations
are of course temperature dependent and obey $v_s(T_c)=0$, $b(T_c)=B(T_c)$ and
$C(T_c)=\Gamma(T_c)$.

We will investigate the stability of the superflow via a QNM analysis of the $U(2)$
model. This automatically will give new and valuable information about the
usual $U(1)$ holographic superfluid since a subsector of the linear fluctuations in
the $U(2)$ model is isomorphic to it.

In section two we will follow \cite{Herzog:2008he,Basu:2008st} and reproduce
the phase diagram based on the comparison of the free energy of the superflow with
the normal phase. Then we will study the QNM spectrum with the superflow. 
In particular we will calculate the direction dependent speed of sound. We will
indeed find that as the superfluid velocity is increased the speed of sound in
opposite direction to the superflow is diminished and eventually vanishes at a
critical velocity $v_{c}$. Increasing the superfluid velocity even further this sound
velocity becomes negative and this has to be interpreted as the appearance of a
negative energy state in the spectrum. In principle that would be enough to
argue for instability but at basically no price the QNM analysis can give us an
even clearer sign of instability. It is well-known that the imaginary part of
the QNMs have to lie all in the lower half plane. If they fail to do so an
exponentially growing mode with amplitude $\phi \propto \exp(\Gamma t)$ appears
in the spectrum. It is not necessary for this mode to have zero momentum. In
fact we will see that if we increase the superfluid velocity beyond the critical value
the imaginary part of the sound mode quasinormal frequency moves into the upper
half plane. And it does so attaining a maximum for non-zero momentum.
We will see that this behavior is necessary to connect the phase diagram
continuously to the normal phase. 
Then moving slightly aside we will study the conductivities with superflow. This
has been done before but only in the transverse sector and here we present
results for the longitudinal sector.

Finally we will briefly investigate the fate of the type II Goldstone mode in
the $U(2)$ model. We will study both the gauged and the ungauged model of
\cite{Amado:2013xya}. 
Landau's criterion suggests that these setups do not sustain
any finite superflow since $\min \frac{\epsilon(p)}{p} = 0$ for quadratic
dispersion relations. Again we can not only look at the real part but also at
the imaginary part. We will indeed find poles in the upper half plane for
non-zero momenta for all temperatures and superfluid velocities for the gauged
and the ungauged model \footnote{Models with one $U(1)$ gauge field and 
two complex scalars similar to our ungauged model were studied
before in \cite{Basu:2010fa} and recently in \cite{Cai:2013wma} (see also \cite{Musso:2013ija}).
There the two scalars had however
different masses and this should prevent the appearance of the ungapped
type II Goldstone mode.}.

Let us also mention some shortcomings of our analysis. We always work in the
so-called decoupling limit in which the metric fluctuations are suppressed.
Therefore we do not see the pattern of first and second (and fourth) sound
typical for superfluids. In the decoupling limit only the fourth sound, the
fluctuations of the condensate, survive. Another shortcoming is that we can
apply the Landau criterion only to the QNMs. As in superfluid Helium
there exist most likely other excitations, such as vortices, that might modify the
value of the critical velocity. The question of if and how solitons of
holographic superfluids determine the critical superfluid velocity has been
investigated in \cite{Keranen:2010sx}. 

It is interesting to compare our results to 
the direction dependence of the sound velocities
in a weakly coupled model like the one recently studied in \cite{Alford:2012vn}.

\section{The $U(2)$ superfluid with superflow}

Consider the bulk Lagrangian for a complex scalar field in the fundamental
representation of a $U(2)$ gauge symmetry \cite{Amado:2013xya,Krikun:2012yj}
\begin{equation} \label{eq:Lagrmodel}
S=\int  d^4 x \sqrt{-g}\mathcal{L}=\int d^4 x \sqrt{-g}\left(-\frac{1}{4}F^{\mu\nu
c}F_{\mu\nu}^c-m^2\psi^\dagger \psi-(D^\mu\psi)^\dagger D_\mu\psi\right)\,,
\end{equation}
where
\begin{align}
\psi=
\sqrt{2} \begin{pmatrix}
\lambda \\
\Psi
\end{pmatrix}\,,  \hspace{2cm}A_{\mu}=A_{\mu}^{c}T_c\,,
\hspace{2cm}
D_{\mu}=\partial_{\mu}-iA_{\mu}\,,
\end{align}
where we include the $\sqrt{2}$ in the definition of the scalar field to agree with the equations of \cite{Herzog:2008he}. Following \cite{Hartnoll:2008vx} we choose the mass of the scalar field to be $m^2=-2/L^2$. 
We take the generators of $U(2)$ to be
 \begin{align}
 T_0=\frac{1}{2}\mathbb{I}\,,\hspace{2cm}T_i=\frac{1}{2}\sigma_i\,.
 \end{align}

Since we will work in the probe approximation we do not include the metric
in the dynamical degrees of freedom but simply consider (\ref{eq:Lagrmodel}) 
in the background metric of the Schwarzschild-AdS black brane
\begin{align}
\nonumber ds^2&=-f(r)dt^2 +\frac{dr^2}{f(r)} + \frac{r^2}{L^2} (dx^2+dy^2)\,, \\
f(r)&= \frac{r^2}{L^2} - \frac{M}{r}\,.
\end{align}
The horizon is located at $r_H=M^{1/3} L^{2/3}$ and its Hawing temperature is
$T= 3 r_H/ 4\pi L^2$. By suitable rescalings we can set $L=r_H=1$ and work with
dimensionless coordinates. 

In order to find background solutions corresponding to a condensate with non-vanishing
superfluid velocity we proceed as follows. First note that the scalar field $\lambda(r)$
can be set to zero by a $U(2)$ gauge transformation. For the scalar $\Psi$ we demand then
that the non-normalizable mode vanishes. By a residual $U(1)$ gauge transformation we can also
take $\Psi$ to be real. 

Now we need to define what we mean by the superflow. Let us discuss this for a moment
from a field theory perspective.
In a multi-component superfluid with $U(2)$ symmetry we can in principle construct the four
(super) currents
\begin{equation}\label{eq:currents}
 J^\mu_a = \Phi^\dagger T_a \nabla^\mu \Phi - \left(\nabla^\mu\Phi\right)^\dagger T_a  \Phi\,,
\end{equation}
where $\nabla^\mu = \partial^\mu - i A_a^\mu T_a$ is the covariant derivative and $\Phi$ is
the condensate wave function which transforms as a doublet under $U(2)$. If the condensate is such that 
one of the spatial currents does not vanish we can speak of a state with non-vanishing superflow. By a gauge transformation we can always
assume the condensate to take some standard form, e.g. $\Phi = (0,\phi)^T$ and represent the
non-vanishing superflow in terms of constant gauge fields. Since we are interested in the case
where we break the $U(2)$ symmetry spontaneously to $U(1)$ we will only allow a non-zero
gauge field in the overall $U(1)$ corresponding to the generator $T_0$. Furthermore by 
an $SO(3)$ rotation we can take the gauge field to point into the $x$ direction. From 
(\ref{eq:currents}) it is easy to see that such a superflow has non-vanishing currents 
$J_x^{(0)}$ and $J_x^{(3)}$. In order to find solutions with non-trivial charge we also
need to introduce a chemical potential. Again in order to preserve the full $U(2)$ symmetry
we also allow a chemical potential only for the overall $U(1)$ charge. 

Returning now to Holography these considerations determine the ansatz for the gauge fields to be of the form
\begin{align}
A^{(0)}  = A^{(0)}_t(r) dt + A^{(0)}_x(r) dx\,, \hspace{2cm}  
A^{(3)} = A^{(3)}_t (r) dt + A^{(3)}_x(r) dx\,.
\end{align}

While we introduce sources only for $A^{(0)}$ the fact that also the current
$J_\mu^{(3)}$ is nonvanishing demands that $A^{(3)} \ne 0$.
The physical interpretation for this fact is that the system forces the
appearance of a charge density $\rho^{(3)} \ne 0$ (as noticed already in \cite{Amado:2013xya})
and a current $J_x^{(3)}$ in the vacuum with superflow. 
This is in turn closely related to the presence
of type II Goldstone bosons in the spectrum  \cite{Watanabe:2011ec}.


At this point it is important to note that the above identification is only
valid in the superfluid phase, that is, whenever $\Psi \ne 0$. A constant
background value of the gauge field $A_x$ in the normal phase is not physically
meaningful since there is no notion of superflow.

For the reasons outlined above we choose the asymptotic boundary conditions for the gauge fields to be
\begin{eqnarray}
&&A_t^{(0)} (r\to\infty)= 2 \bar \mu\,,\hspace{2cm}
A_t^{(3)}(r\to\infty)=0\,,\nonumber\\
&&A_x^{(0)}(r\to\infty)=2  \bar S_x \,,\hspace{1.81cm} A_x^{(3)}(r\to\infty)= 0\,,
\end{eqnarray}
where $\bar\mu$ is to be identified with the chemical potential of the dual
theory and $\bar S_x$ is related to the superflow velocity. We have
included a factor of two in the definitions of $\bar\mu$ and $\bar S_x$ for the following 
reason.
The background field equations can be recast in the form of those derived from the
$U(1)$ model in \cite{Herzog:2008he,Basu:2008st} by using the  field redefinitions
\begin{align} \label{eq:lincombin}
A_0=\frac{1}{2}(A^{(0)}_t- A^{(3)}_t)\,, \hspace{2cm}
\xi=\frac{1}{2}(A^{(0)}_t+ A^{(3)}_t)\,, \nonumber\\
A_x=\frac{1}{2}(A^{(0)}_x- A^{(3)}_x)\,, \hspace{2cm}
\varsigma=\frac{1}{2}(A^{(0)}_x+ A^{(3)}_x)\,,
\end{align}
for which the background equations now read
\begin{align}
\label{eq:psieq}\Psi'' + \left(\frac{f'}{f} + \frac{2}{r}\right) \Psi' + \left(
\frac{A_0^2}{f^2}- \frac{A_x^2}{r^2 f} -\frac{m^2}{f}\right)\Psi =0\,,\\
\label{eq:chieq} A_0'' + \frac{2}{r} A_0' -\frac{2 \Psi^2}{f}A_0=0\,,\\
\label{eq:Axeq} A_x'' + \frac{f'}{f}A_x'- A_x \frac{2 \Psi^2}{f}=0\,,\\
\label{eq:xieq} \xi''+\frac{2}{r}\xi' = 0\,,\\
\label{eq:varsigmaeq} \varsigma'' + \frac{f'}{f}\varsigma' = 0\,.
\end{align}
It can be checked that
we recover the usual $U(1)$ system  describing the $U(1)$ holographic superconductor in the
presence of superfluid velocity (see for instance \cite{Arean:2010zw}).
The chemical potential $\bar \mu$ is therefore the chemical potential for the field
$A_0$ which plays the role of the temporal component of the (single) gauge field, 
and $A_x$ plays the role of the spatial component of the single gauge field of \cite{Herzog:2008he,Basu:2008st,Arean:2010zw}. 
This explicitly shows that the background of the $U(2)$ model is identical to that of
the $U(1)$ superconductor, even for a nonzero superfluid velocity. 

An immediate consequence of the fact that the background equations are those of
the $U(1)$ holographic superfluid is that, at first sight, the $U(2)$ system
seems to be able to accommodate a superflow. However, as already argued, this is
in direct contradiction with the Landau criterion of superfluidity
\cite{PinesNozieres} due to the presence of a type II Goldstone in the spectrum.
Of course, having found solutions to the equations of motion does not yet
say anything about the stability. In fact as we will explicitly see the
type II Goldstone will turn into an unstable mode and therefore make the whole
$U(2)$ solution with superflow unstable.

Equations (\ref{eq:psieq})-(\ref{eq:Axeq}) are non-linear and have to be solved
using numerical methods. Notice that (\ref{eq:xieq}) and
(\ref{eq:varsigmaeq}) are decoupled. They correspond to the preserved $U(1)$
symmetry after having broken spontaneously $U(2) \rightarrow U(1)$. The
asymptotic behavior of the fields close to the conformal boundary is
\begin{eqnarray}
&&A_0= \bar \mu - \frac{\bar \rho}{r}+\dots\,,\nonumber\\
&&\label{eq:bdcond}A_x=\bar S_x - \frac{\mathcal{\bar J}_x}{r} + \dots
\,,\\
&&\Psi= \frac{\psi_1}{r} + \frac{\psi_2}{r^2}+\dots\,.\nonumber
\end{eqnarray}
The asymptotic quantities are related to the physical ones by
\begin{eqnarray}
&&\bar \mu = \frac{3 }{4 \pi T}\mu\,,\quad\quad\quad
\bar \rho = \frac{9}{16 \pi^2 T^2} \rho\,,\nonumber\\
&&\bar S_x = \frac{3 }{4 \pi T}S_x\,,\qquad 
\mathcal{\bar J}_x =\frac{9}{16 \pi^2 T^2}\mathcal{J}_x\,,\\
&&\psi_1 =  \frac{3}{4 \pi T} \left< \mathcal{O}_1 \right>\,,\quad
\psi_2 =  \frac{9}{16 \pi^2 T^2} \left< \mathcal{O}_2 \right>\,.\nonumber
\end{eqnarray}

We are working in the grand canonical ensemble, then we fix the
chemical potential $\mu$. The temperature is defined by  $T/\mu \propto 1/\bar
\mu$. For studying the evolution of the condensate as a function of the
superfluid velocity, the natural way to proceed is to work with $S_x/\mu$ as our
free parameter together with temperature. Notice that both asymptotic modes of
the scalar field are actually normalizable \cite{Klebanov:1999tb}. From now on
we will stick to the $\mathcal{O}_2$ theory, for which $\psi_1=0$ and
$\left<{\mathcal{O}_2}\right>$ is the vev of a scalar operator of mass dimension
two in the dual field theory. Notice that the fields $\xi$ and $\zeta$ corresponding to the unbroken $U(1)$ are given by 
\begin{align}
 \xi&= \bar\mu - \bar\rho/r\,,\nonumber\\
 \zeta &= \bar S_x\,,
\end{align}
even with non-vanishing condensate.

\begin{figure}[htp!] 
\centering
\includegraphics[width=300pt]{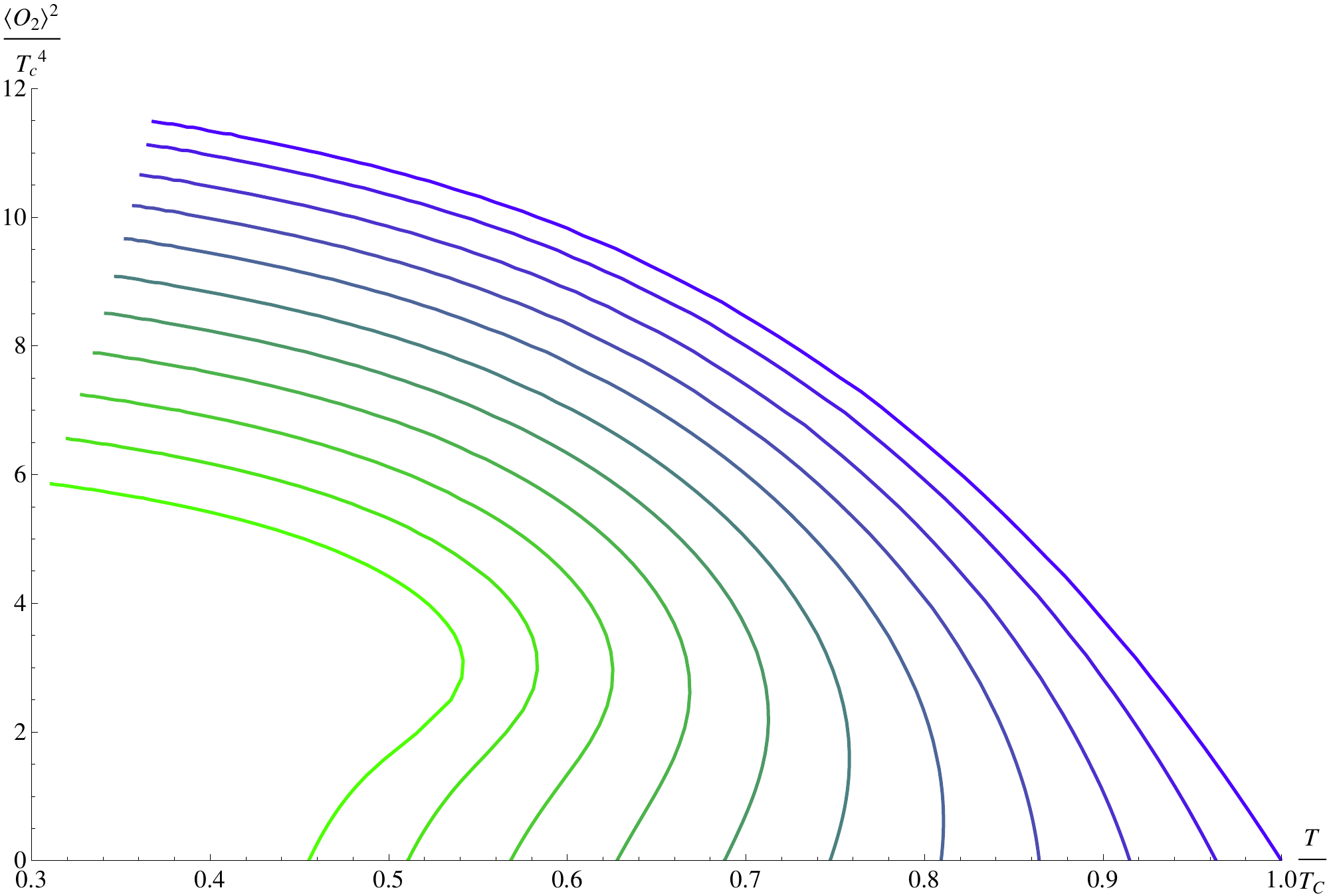}
\caption{The condensate for different values of the superfluid velocity, ranging from
$\frac{S_x}{\mu}=0.005$ (right) to $\frac{S_x}{\mu}=0.530$ (left).  }
\label{fig:Background}
\end{figure}

The values of the condensate as a function of temperature and superfluid
velocity shown in Figure \ref{fig:Background} reproduce the previous results of
\cite{Herzog:2008he,Basu:2008st}. In the plot and in the rest of the paper the
temperature is measured with respect to the critical temperature of the phase
transition with no superfluid velocity, i.e. $T_c \approx 0.0587 \mu$. 

\subsection{Free Energy}

In this section we compute the free energy of the condensed phase and compare it
to the free energy of the unbroken phase as done in \cite{Herzog:2008he,Basu:2008st}. 
After
appropriate renormalization of the Euclidean on-shell action and using the
boundary conditions ({\ref{eq:bdcond}), the free energy density reads
\begin{equation}
F= -T S_{ren} = -\bar\mu\bar\rho + \bar{S_x}\mathcal{\bar J}_x +\int_1^\infty dr
\left(\frac{2 r^2 A_0^2}{f}-2A_x^2\right) \Psi^2\,.
\end{equation}
In the normal phase $\Psi=0$, regularity at the horizon forces the $A_x$ gauge
field to have a trivial profile along the radial direction in the bulk and
therefore not to contribute to the free energy, i.e. $\mathcal{\bar J}_x=0$. This is in
accordance with the fact that in absence of a scalar condensate it is not
possible to switch on a superfluid velocity anymore. Switching on the spatial
component of the gauge field in the normal phase describes a pure gauge
transformation that does not affect the free energy of the system. In the broken
phase instead, different superfluid velocities are physically distinguishable.
It is important to emphasize that one is actually comparing the normal
phase at vanishing superfluid velocity with the superconducting phase at different
values of the superfluid velocity, and that the normal phase is unstable towards
condensation without superflow for any $T<T_c$. Therefore, the physical
relevance of this comparison is not completely clear. We will see later on that
actually the Landau criterion establishes a different transition temperature for
the superfluid phase. Nevertheless the free energy gives a natural first approach
to characterize the phase diagram of the system.

\begin{figure}[htp!]
\centering
\includegraphics[width=225pt]{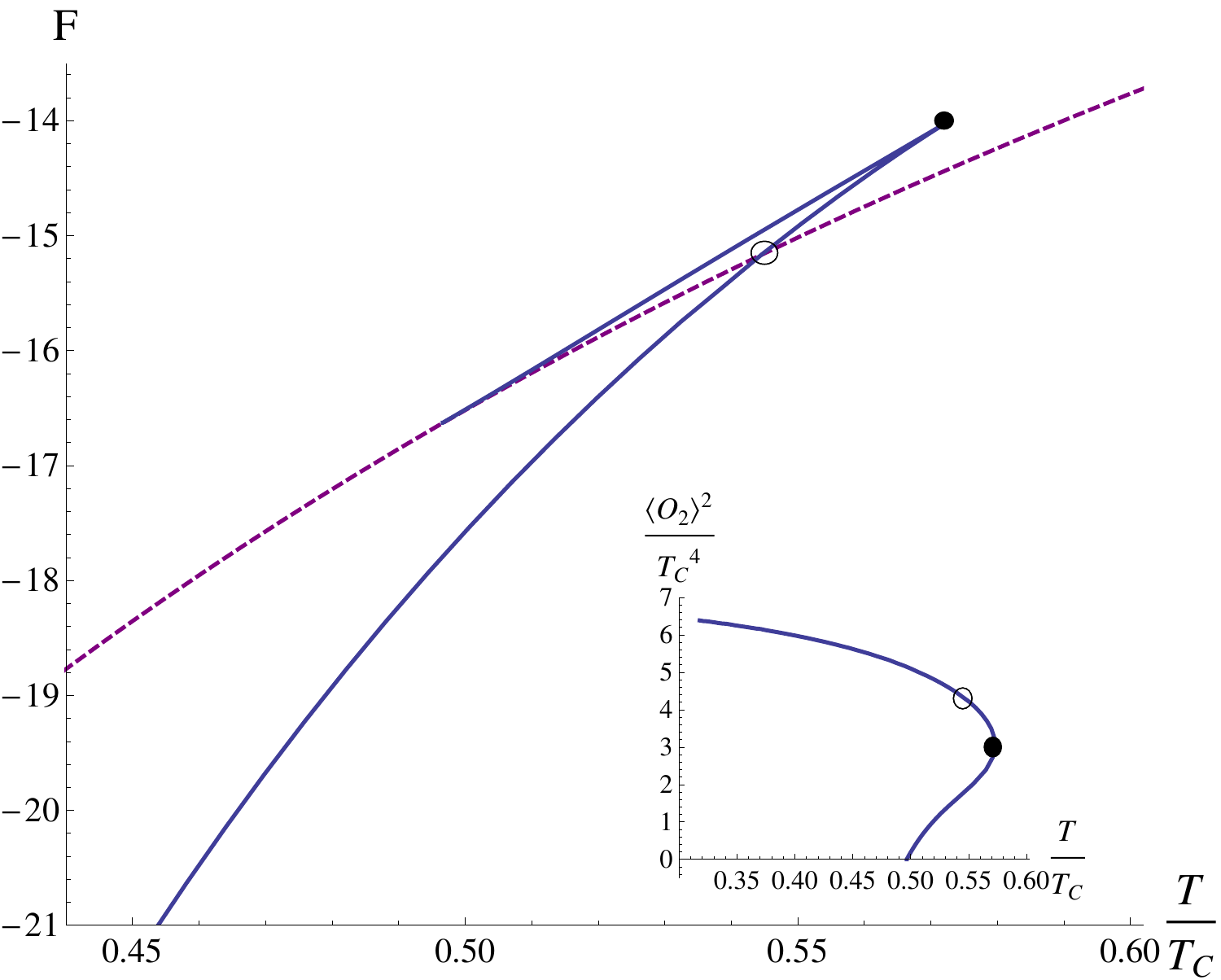} \hfill
\includegraphics[width=225pt]{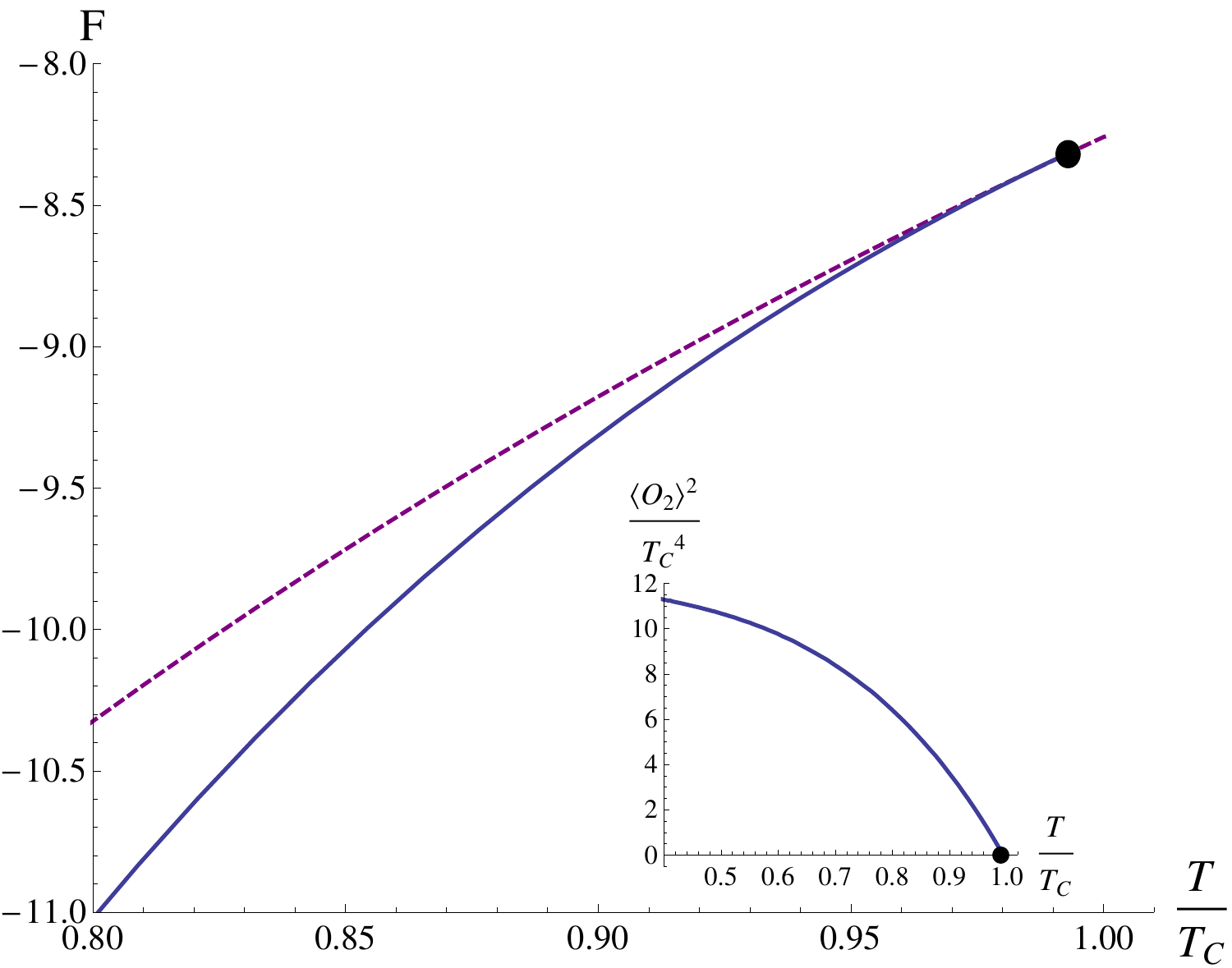}
\caption{Free energy of the condensed (solid line) and normal (dashed line)
phases for $\frac{S_x}{\mu}=0.5$ (left) and $\frac{S_x}{\mu}=0.05$ (right). The
small plots show the behavior of the condensate. The open circle corresponds
to the critical temperature $\tilde{T}$ whereas the filled circle corresponds to the
spinodal point (max. overheating).}
 \label{Fe1}
\end{figure}

In Figure \ref{Fe1} the free energy of both the normal and condensate phase
is plotted for different values of $\frac{Sx}{\mu}$. The different behavior
for large and small values of the superfluid velocity is apparent. 
For large superfluid velocity the transition is first order as can be seen
from the left panel in Figure \ref{Fe1}, indicated by the open circle. 
Coming from low temperatures the
system can still be overheated into a metastable state until the point of spinodal decomposition
where the order parameter susceptibility $\partial \langle \mathcal{O} \rangle / \partial \mu$
diverges, indicated by the filled circle. 

For low superfluid velocities the normal phase free energy and the condensate free
energy match smoothly at a second order phase transition. 
The resulting phase space is contained in Figure
\ref{fig:phasespace} and  reproduces the previous analysis in \cite{Herzog:2008he,Basu:2008st}. 

The phase transition found from considerations of the free energy
is however only apparent. We will call the temperature at which the
free energies of the condensate phase with superflow and the free energy
of the normal phase coincide $\tilde T$ from now on. The temperature
at which the (second order) phase transition occurs without superflow we
will denote by $T_c$. As we will show now the superflow becomes unstable
at temperatures below $\tilde T$ as implied by the Landau criterion applied
to the sound mode. This temperature we will denote by $T^*$.

									

\section{Landau criterion for the $U(1)$ sector}

In this section we analyze the QNM spectrum of the $(0)-(3)$ 
sector, which is identical to the original $U(1)$ holographic superconductor in
the presence of superfluid velocity \cite{Herzog:2008he,Basu:2008st}. We focus on the
behavior of the lowest QNM, the type I Goldstone boson, with
special emphasis on the velocity and the attenuation constant and their
dependence on the superfluid velocity and on the angle of propagation with respect
to the flow.

To study the QNM spectrum we consider linearized perturbations around the background of the fields 
of the form $\delta \phi_I = \delta \phi_I(r) \exp[-i(\omega\, t-|k|\,x\cos(\gamma) - |k|\, y \sin(\gamma)]$. Specifically we consider the fluctuations
\begin{eqnarray}
\label{eq:pertsc} \delta \hat \Psi^T &=& (\eta(r), \sigma(r)) \,,\nonumber\\
\label{eq:perta0} \delta A^{(0)} &=& a^{(0)}_t(r)dt +  a^{(0)}_x(r)dx +  a^{(0)}_y(r)dy\,,\\
\label{eq:perta3} \delta A^{(3)} &=& a^{(3)}_t(r)dt +  a^{(3)}_x(r)dx +  a^{(3)}_y(r)dy\,,\nonumber
\end{eqnarray}
where in the case of the gauge fluctuations we will work with the linear combinations already defined by (\ref{eq:lincombin}), i.e. $a^{(-)}_\mu \equiv \frac{1}{2}( a^{(0)}_\mu - a^{(3)}_\mu)$ and
$a^{(+)}_\mu \equiv \frac{1}{2}( a^{(0)}_\mu + a^{(3)}_\mu)$.
The linearized equations are rather complicated and we list them in Appendix A. The numerical techniques used
to obtain the hydrodynamic modes in coupled systems are well known. We will not
elaborate on them here, referring the interested reader to \cite{Amado:2009ts}
and \cite{Kaminski:2009dh}.

\begin{figure}[ht] 
\centering
\includegraphics[width=225pt]{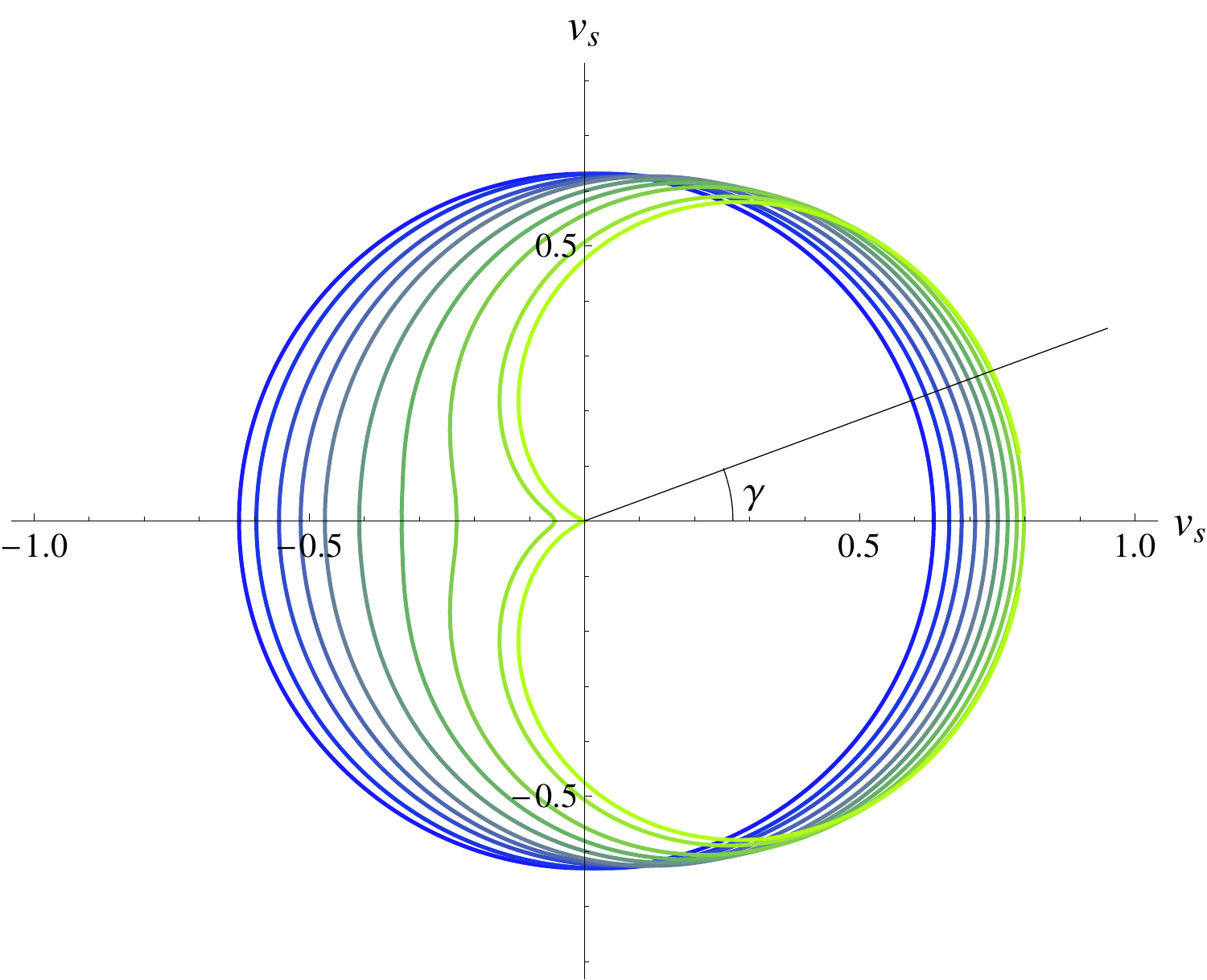}
\hfill \includegraphics[width=225pt]{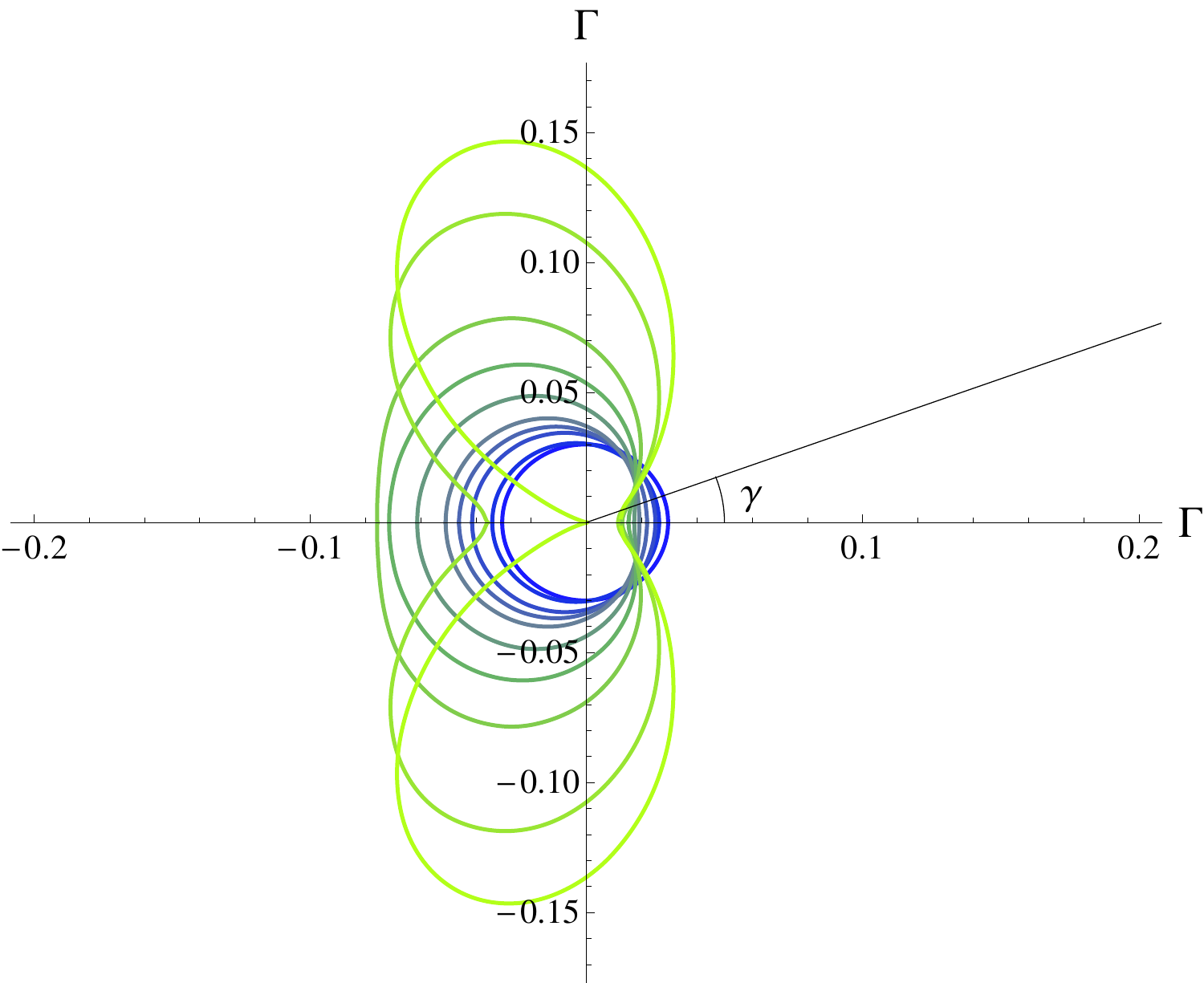}
\caption{Sound velocity and damping for $T=0.7T_c$ and several superfluid velocities 
from $S_x/\mu = 0$ (blue) to $S_x/\mu =0.325$ (green). The radius represents the
absolute value of the sound velocity (left) and attenuation constant (right) as
a function of the angle $\gamma$ between the momentum and the superfluid velocity.}
\label{fig:pica0}
\end{figure}

\begin{figure}[htp!] 
\centering
\includegraphics[width=225pt]{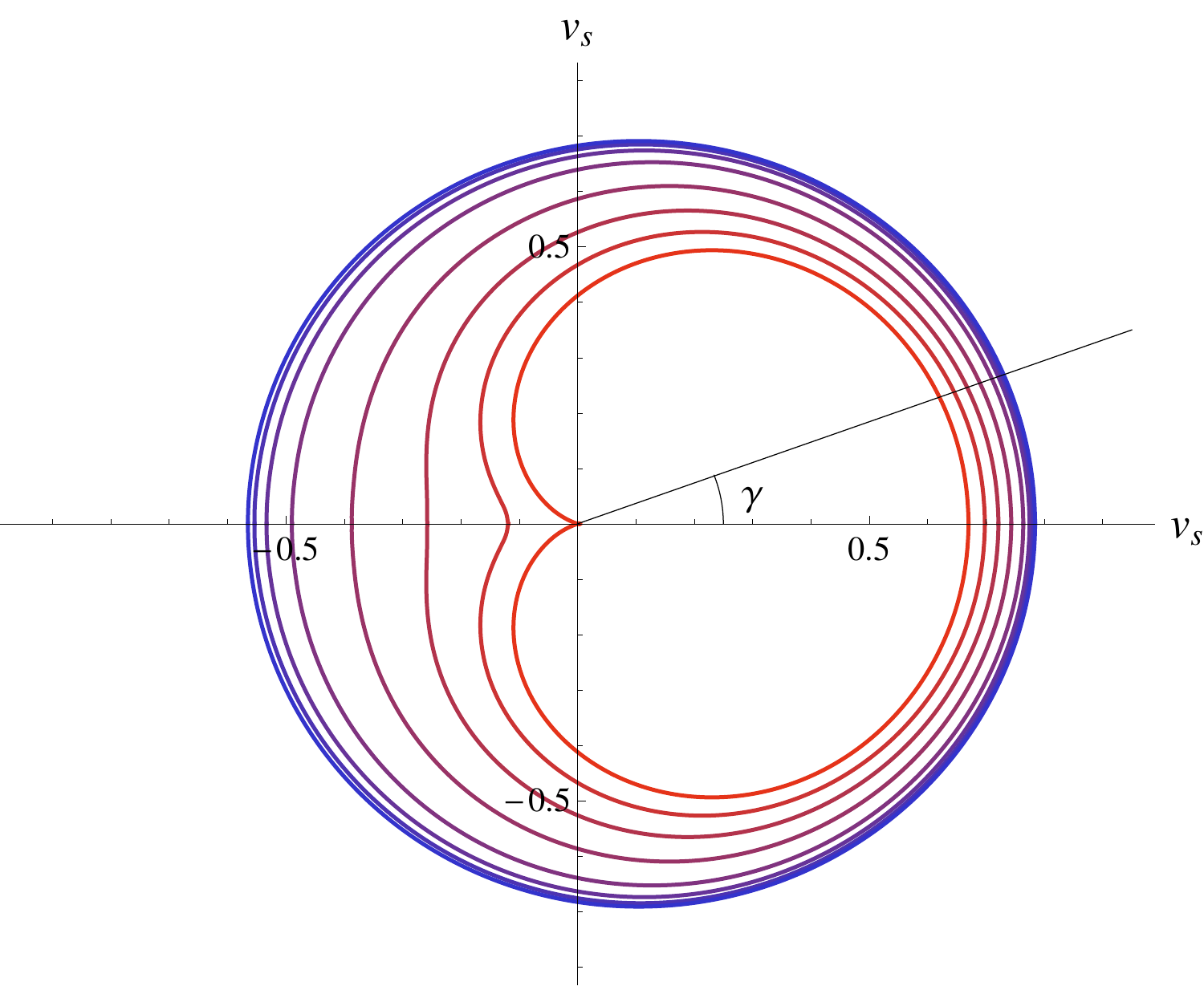}
\hfill \includegraphics[width=225pt]{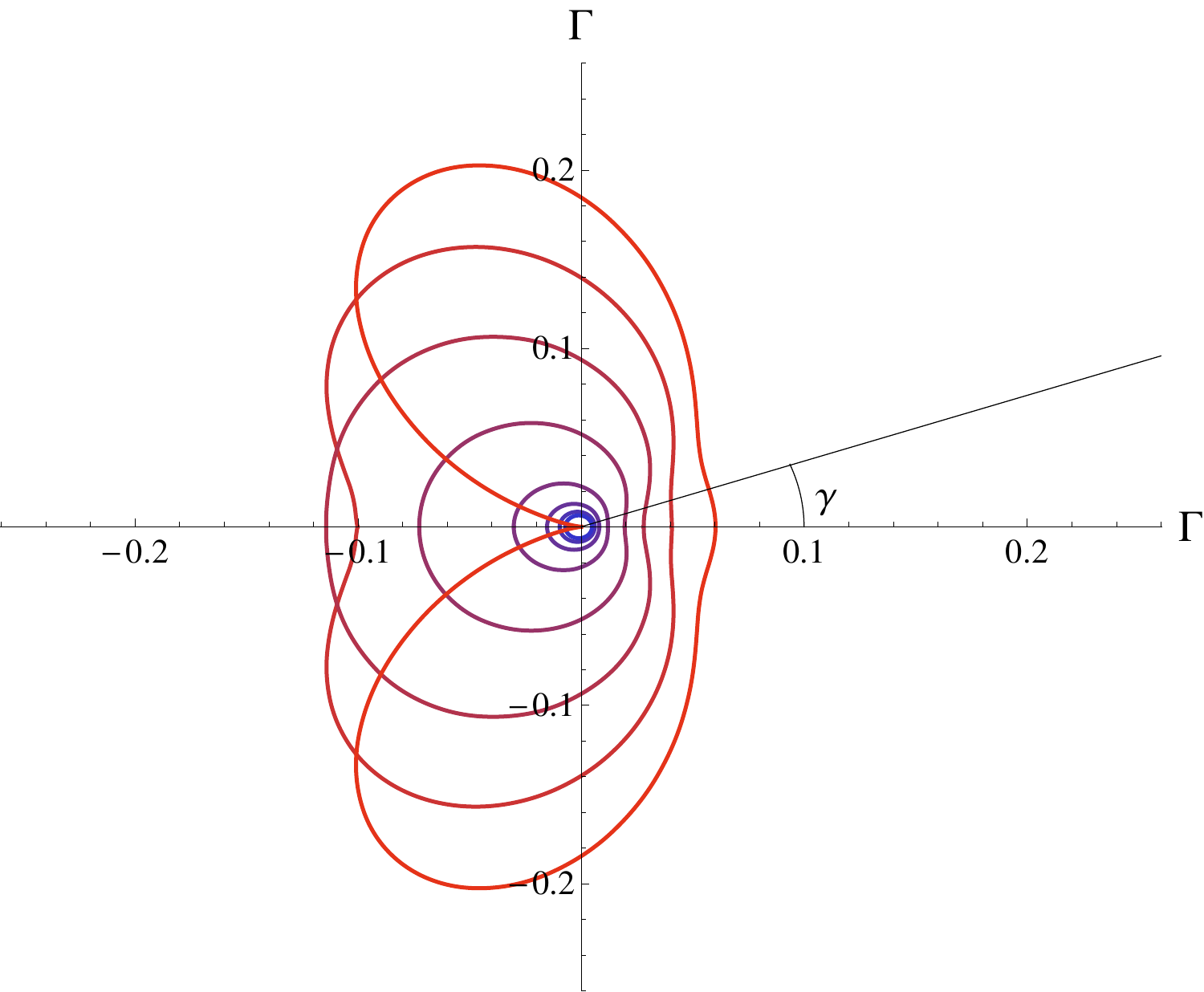}
\caption{Sound velocity (left) and attenuation constant (right) for $S_x/\mu =
0.2$ as a function of the angle $\gamma$ and for a range of temperatures from $T=0.85T_c$ (red) to
$T=0.57T_c$ (blue).}
\label{fig:pica1}
\end{figure}

In Figures \ref{fig:pica0} and \ref{fig:pica1} we represent the velocity and the
attenuation of the type I Goldstone mode. Its dispersion relation is given by
(\ref{eq:sound}) at low momentum, except now the speed of sound $v_s$
and the attenuation constant $\Gamma$ depend on the angle $\gamma$
\footnote{The small real constant $b$ does not play a role here since for small enough 
momentum the linear part proportional to $v_s$ dominates.}.
Figure \ref{fig:pica0} shows the angle dependent variation of the sound velocity and damping
constant for a fixed temperature and varying values of the
superfluid velocity. Figure \ref{fig:pica1} shows the same at fixed
superfluid velocity but with varying temperature. As one would expect for small $S_x/\mu$ and low enough temperature the velocity and damping constant are almost isotropic. 
As the superfluid velocity is
increased or the temperature is increased 
the plot becomes more and more asymmetric. The anisotropy of
the system is such that we see an enhancement of the sound velocity and a
reduction of the damping in the direction of the superflow. 

\begin{figure}[ht] 
\centering
\includegraphics[width=225pt]{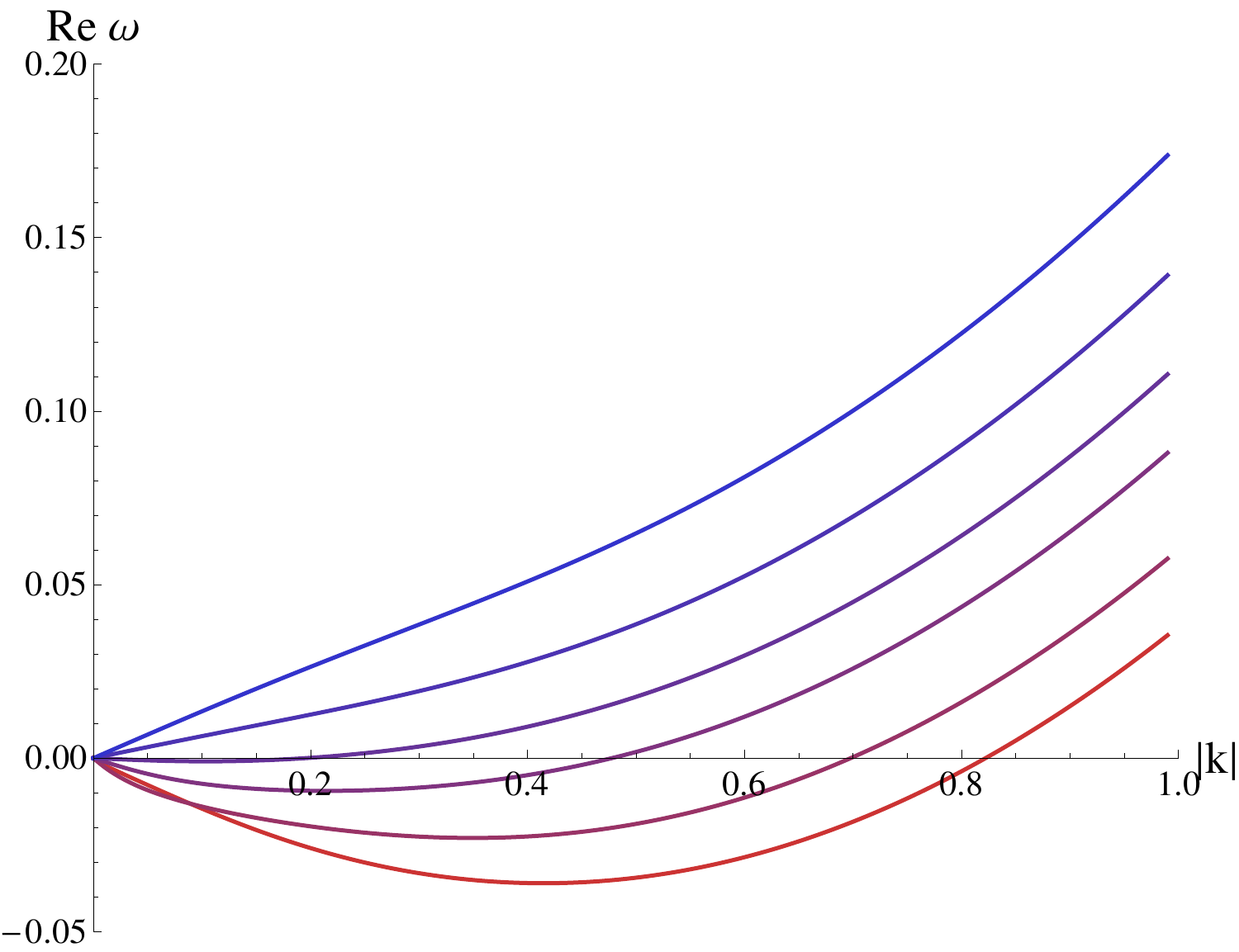}
\hfill \includegraphics[width=225pt]{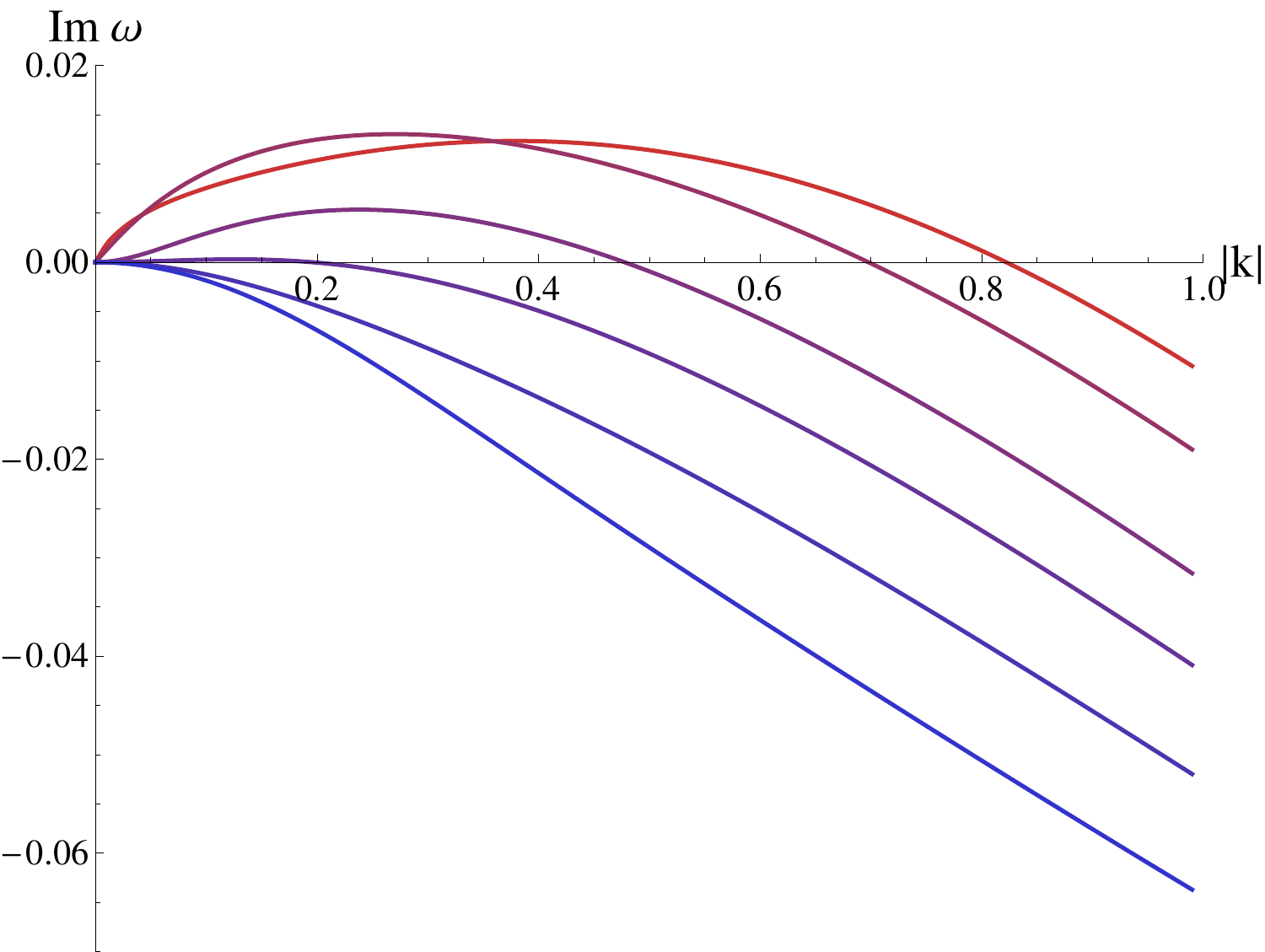}
\caption{Real (left) and imaginary (right) parts of the frequency of the lowest
hydrodynamic mode (type I Goldstone mode) versus momentum at $S_x/\mu= 0.1$ and
$\gamma=\pi$ for different temperatures from $T=\tilde{T}=0.970T_c$ (red) to
$T=0.905 Tc$ (blue). The instability appears at $T^*=0.935T_c$.}
\label{fig:TachQNMstriped}
\end{figure}

\begin{figure}[ht] 
\centering
\includegraphics[width=225pt]{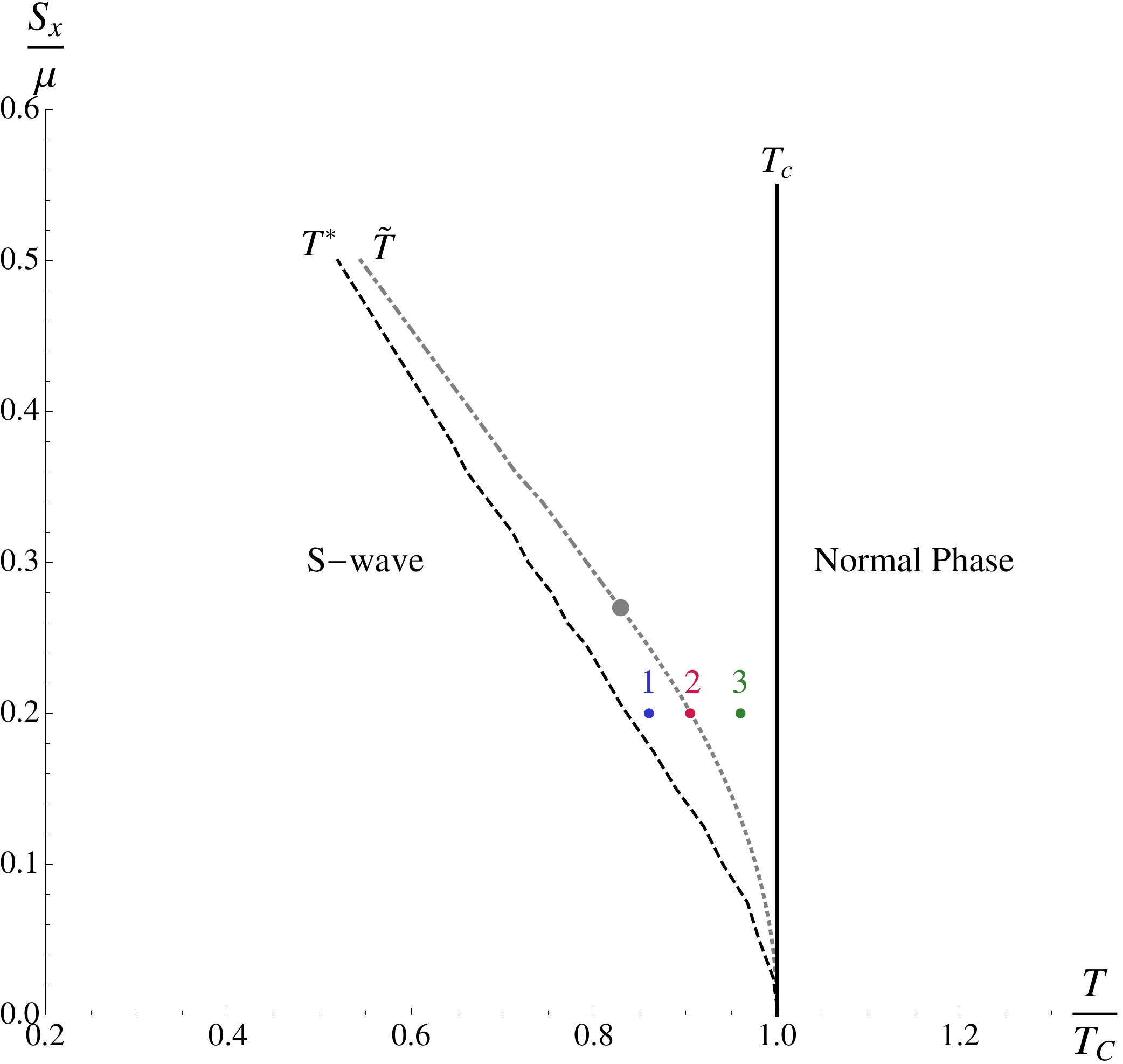}
\hfill \includegraphics[width=225pt]{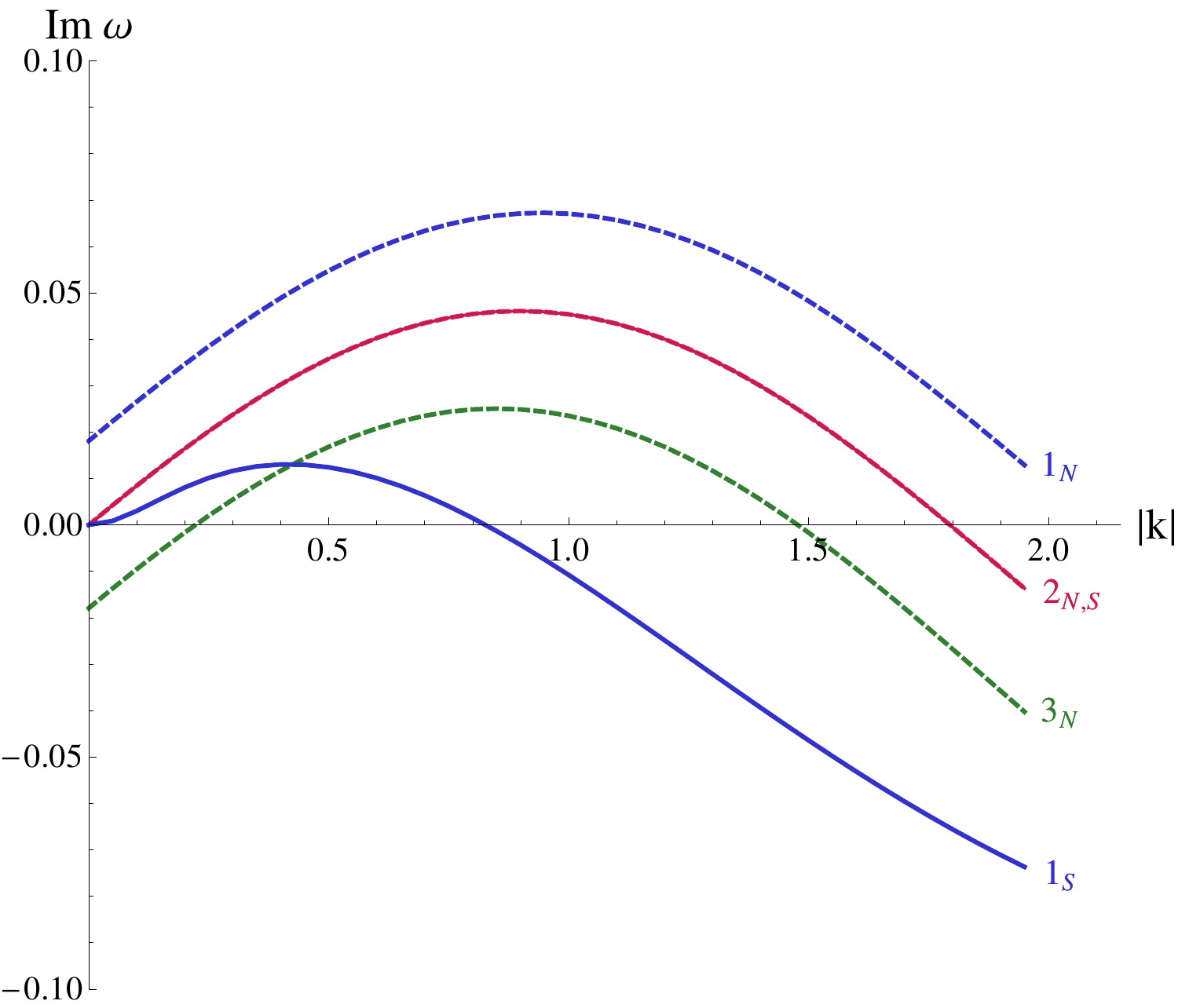}
\caption{(Left) Phase diagram after the study of the QNMs. The grey dashed
line corresponds to $\Tilde{T}$, the apparent transition temperature found by direct
analysis of the free energy. At a certain point (disk) the transition in free energy changes from
2nd order (dotted) to 1st order (dash-dotted). The black solid line corresponds to
the critical temperature in absence of superfluid velocity. The black dashed
line signals the physical phase transition at $T^*$, the temperature at which the
local instability appears. Points 1, 2 and 3 indicate the values of
temperature and velocity used in the plot on the right. (Right) Imaginary part of the
lowest QNM for different temperatures at fixed $S_x/\mu=0.2$ and $\gamma=\pi$.
Dashed lines were obtained in the normal phase whereas solid lines were
calculated in the superfluid phase.}
\label{fig:phasespace}
\end{figure}

The most interesting feature of the system is found however in the opposite 
direction to the superfluid velocity. As one can see in both plots, at $\gamma=\pi$ the
reduction in the sound velocity is strongest and eventually both the
attenuation constant and the sound velocity vanish simultaneously. It is
important to stress that this happens below the temperature $\tilde{T}$. If one
continues increasing the temperature (or equivalently increasing the superfluid velocity at
fixed temperature) one finds that the real part of the frequency becomes
negative and that its imaginary part crosses to the upper half plane, as depicted in
Figure \ref{fig:TachQNMstriped}. This signals the appearance of a tachyonic mode.
$T^*$ is the temperature where both the instability appears and the
speed of sound becomes negative. This temperature actually signals the
end of the superfluid phase according to the Landau criterion, and therefore we
interpret it as the physical phase transition temperature.

In Figure \ref{fig:phasespace} (left) we present the phase diagram resulting from
the QNM analysis. To illustrate the situation, on the right plot we show the behavior
of the relevant QNM\footnote{In the unbroken phase this is just the lowest scalar QNM, while in the
broken phase it is the sound mode at fixed $S_x/\mu$.} at three different points of the
phase diagram (points labelled 1, 2, 3 on the left plot).
At $\tilde T<T<T_c$ in the unbroken phase (line $3_N$), the mode that was
responsible for the transition to the homogeneous superfluid
phase without superfluid velocity is shifted and becomes unstable at finite momentum.
This behavior reflects the fact that the system is unstable for $T\leq T_c$, the mode being
shifted in momentum due to the constant nonzero value of $A_x$.
At $T=\tilde T$ (lines $2_{N,S}$) the lowest mode becomes unstable at $k=0$. It is at this point
that the free energy of the homogeneous superfluid phase equals that of the normal
phase.
Hence, the free energy analysis, which only captures the $k=0$ dynamics, predicts a phase transition at this temperature. For the particular superfluid velocity in the plot the phase transition is second order.
Finally, the fate of the QNM for $T^*<T<\tilde T$ is shown in lines $1_N$ (for the normal
phase) and $1_S$ (for the homogeneous superflow phase). One can see that the
Goldstone mode in the superfluid phase is unstable for a finite range in momentum.
Only at $T^*$ this mode becomes stable again as shown in Figure \ref{fig:TachQNMstriped}.
It is at this temperature that the homogeneous superflow phase becomes stable
according to the Landau criterion since the sound velocity becomes positive (moreover the
imaginary part of the QNM dispersion relation lies entirely in the lower half plane).

Therefore the QNM results indicate that a phase transition occurs at a 
lower temperature $T^* <\tilde{T}$. Similarly, if we imagine the
system at fixed temperature and start rising the superfluid velocity, both $v_s$ and
$\Gamma$ will vanish at some value of $S_x/\mu$, which we claim is indeed the
critical velocity $v_c$ of the superfluid, in the sense of the Landau criterion.

As a very interesting fact, notice that the imaginary part of the mode
exhibiting the instability has a maximum at finite momentum as well. The fact
that the instability appears at finite momentum suggests that there might exist 
a new  (meta)stable intermediate phase above $T^*$ with a spatially
modulated condensate. 
Examples of such instabilities towards spatial modulation 
have been discussed before in \cite{Nakamura:2009tf, Donos:2011bh,
Bayona:2011ab}.

Recall that the Landau criterion is formulated uniquely in terms of
$\Re(\omega)$. At a given temperature the critical velocity corresponds to the
superfluid velocity at which $v_s =0$, or equivalently to the value of $S_x/\mu$
where $\Re(\omega)$ becomes negative (see Figure \ref{fig:TachQNMstriped}). That
the criterion is a statement about $\Re(\omega)$ reflects the fact that it holds
also at zero temperature. At finite temperature the
dispersion relation of the gapless mode gets itself altered due to both the
superfluid velocity and the temperature \cite{PinesNozieres, Alford:2012vn}, implying
that generically the critical value of $S_x/\mu$ at fixed temperature \emph{does
not} correspond to the velocity of sound at the same
temperature and vanishing superfluid velocity.

An extra comment is in order here regarding the phase of the system for
$T_c>T>\tilde{T}$. The fact that in the unbroken phase the lowest QNM is
unstable in this regime (see line $3_{N}$ in Figure \ref{fig:phasespace}) of
course indicates that the normal phase is unstable. Let
us comment on this. Since the condensate vanishes in the normal phase, there
exists no physical notion of superfluid velocity in this phase; different
choices of $A_x$ are just different frame choices. In particular, a constant
$A_x$ simply acts as a shift in momentum in the unbroken phase, as can be seen
from the fact that the maximum of the QNM is centered at a momentum equal to the
value of the gauge field at the conformal boundary. Therefore the normal phase
is unstable for any temperature lower than the critical temperature $T_c$
towards the formation of a superfluid without superflow. On the other hand, we
know that the homogeneous condensate solution with finite velocity does not exist
in this region, and moreover it is unstable for $T>T^*$. 
We see two possibilities for the completion of the phase diagram in this region.
First, the system could simply fall down to the true ground state, which is the
condensate with no superflow. At finite $S_x/\mu$ this is still a solution
which minimizes the energy albeit with a condensate that is not real anymore
but rather has a space dependent phase such that $\vec{\nabla}\Phi=0$.
This is simply the gauge transformed homogeneous ground state without superflow.
On the other hand, the fact that we found an instability at finite momentum
in the temperature range $T^*<T<\tilde T$ 
could indicate that there is a spatially modulated (metastable) phase even
in the range $T^*<T<T_c$, namely a striped
superfluid. Due to the smooth appearance of the unstable mode we expect the transition at
$T^*$ to that phase to be 2nd order, although this should be studied in
detail by constructing the correct inhomogeneous background. 
The explicit construction of this phase goes however substantially beyond the purpose
of this paper and we leave this question open for further investigation.

\subsection{Longitudinal conductivities in the $U(1)$ sector}

In this section we compute the conductivities in the $(0)-(3)$ sector in the
presence of superfluid velocity. As far as we are aware, only the transverse
conductivities have been computed so far (see for instance \cite{Arean:2010zw,Arean:2010xd}).
In contrast, here we will focus on the longitudinal conductivities. These are calculated, via the Kubo formula
\begin{equation}
\sigma = \frac{i}{\omega} \langle J^x J^x \rangle \, ,
\end{equation}
from the two point function
\begin{equation}
 \mathcal{G}_{IJ}=\lim_{\Lambda\rightarrow\infty}\left(\mathcal{A}_{IM}\mathcal{F}^M_{kJ}(\Lambda)'\right)\,,
\end{equation}
where the  matrix  $\mathcal{A}$ can be read off from the on-shell action. $\mathcal{F}$ is the matrix valued bulk-to-boundary propagator normalized to the unit matrix at the boundary. Since we are only interested 
in the entry of the matrix corresponding to $\langle J^x J^x \rangle$ and the matrix $\mathcal{A}$
is diagonal, we just need one element, i.e. $\mathcal{A}_{xx}=-\frac{f(r)}{2}$. In order to construct the bulk-to-boundary propagator one needs  
a complete set of linearly independent solutions for the perturbations of the scalar and gauge fields. This implies 
solving the system of equations given in Appendix A at zero momentum. The method follows closely the one detailed in
\cite{Kaminski:2009dh}.
Notice that there is a surviving coupling between the gauge fields and the scalar 
perturbations mediated by $A_x$. This makes the
computation of the conductivities more involved than in the case without superflow. 

\begin{figure}[ht] 
\centering
\includegraphics[width=225pt]{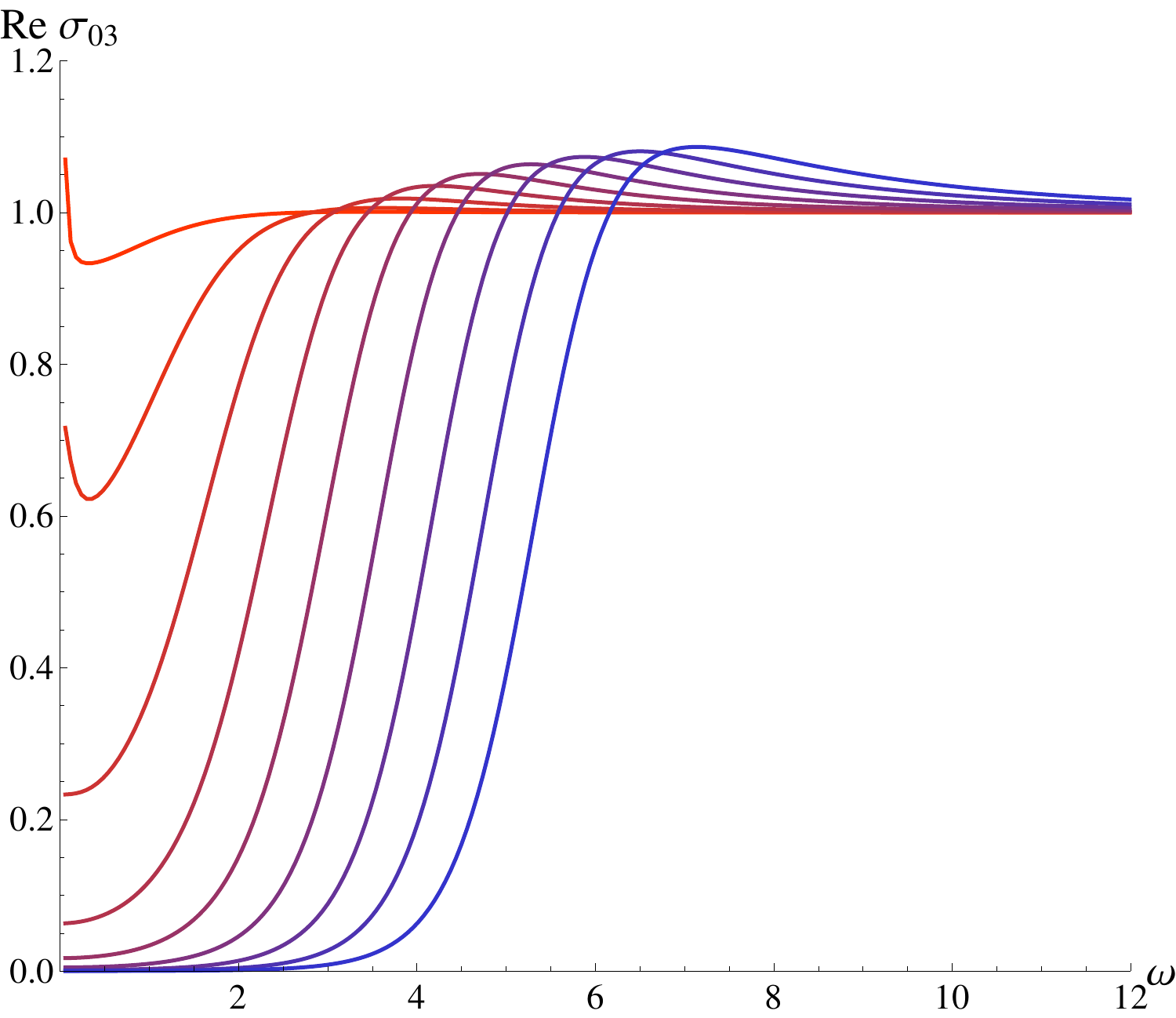}\hfill
\includegraphics[width=225pt]{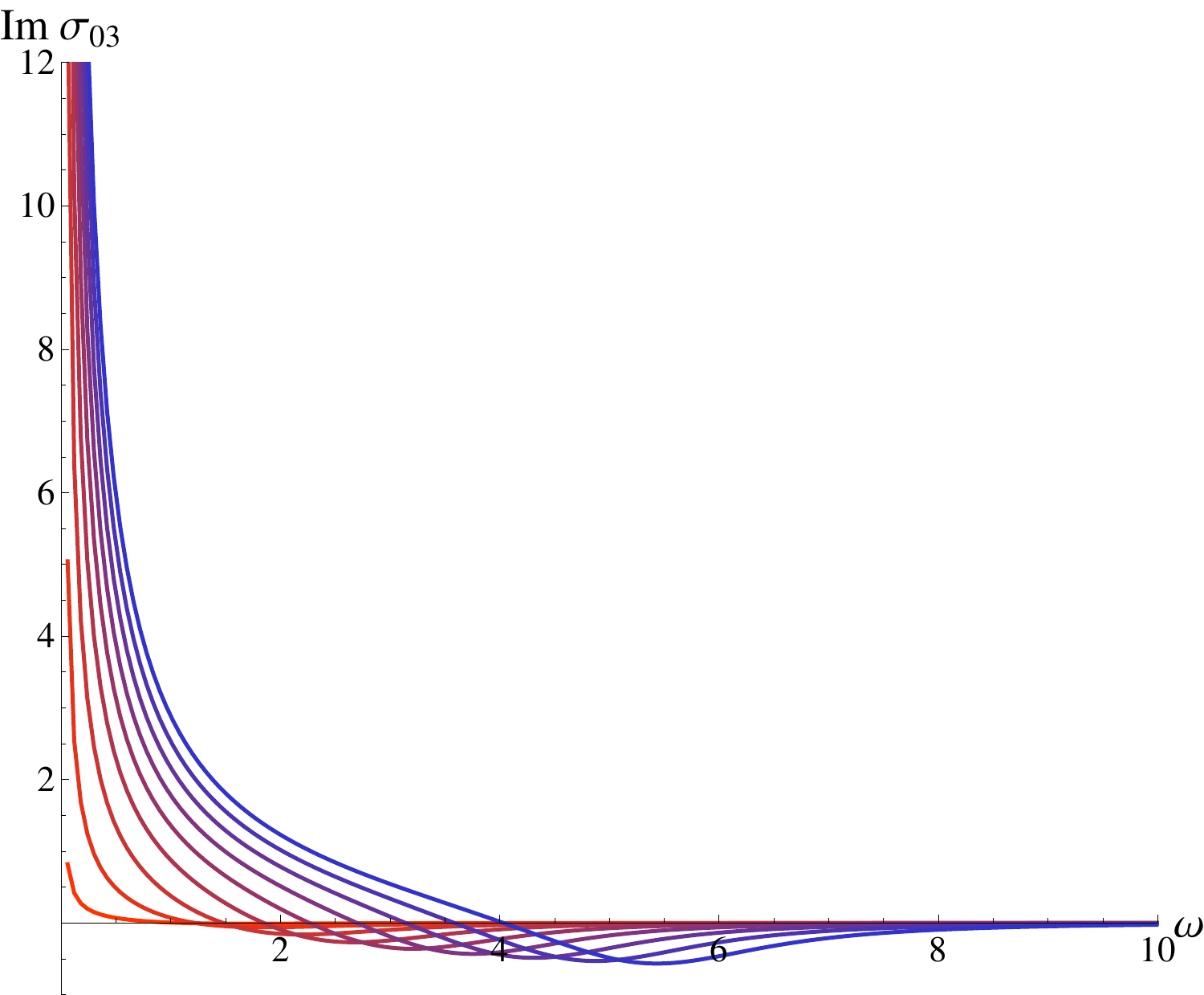}
\caption{Plots of the Real (left) and Imaginary (right) parts of the
conductivity for fixed $S_x/\mu=0.05$. Different lines correspond to different
temperatures from $T=0.99T_c$(red) to $T=0.38T_c$ (blue).}
\label{fig:condlow}
\end{figure}

\begin{figure}[ht] 
\centering
\includegraphics[width=225pt]{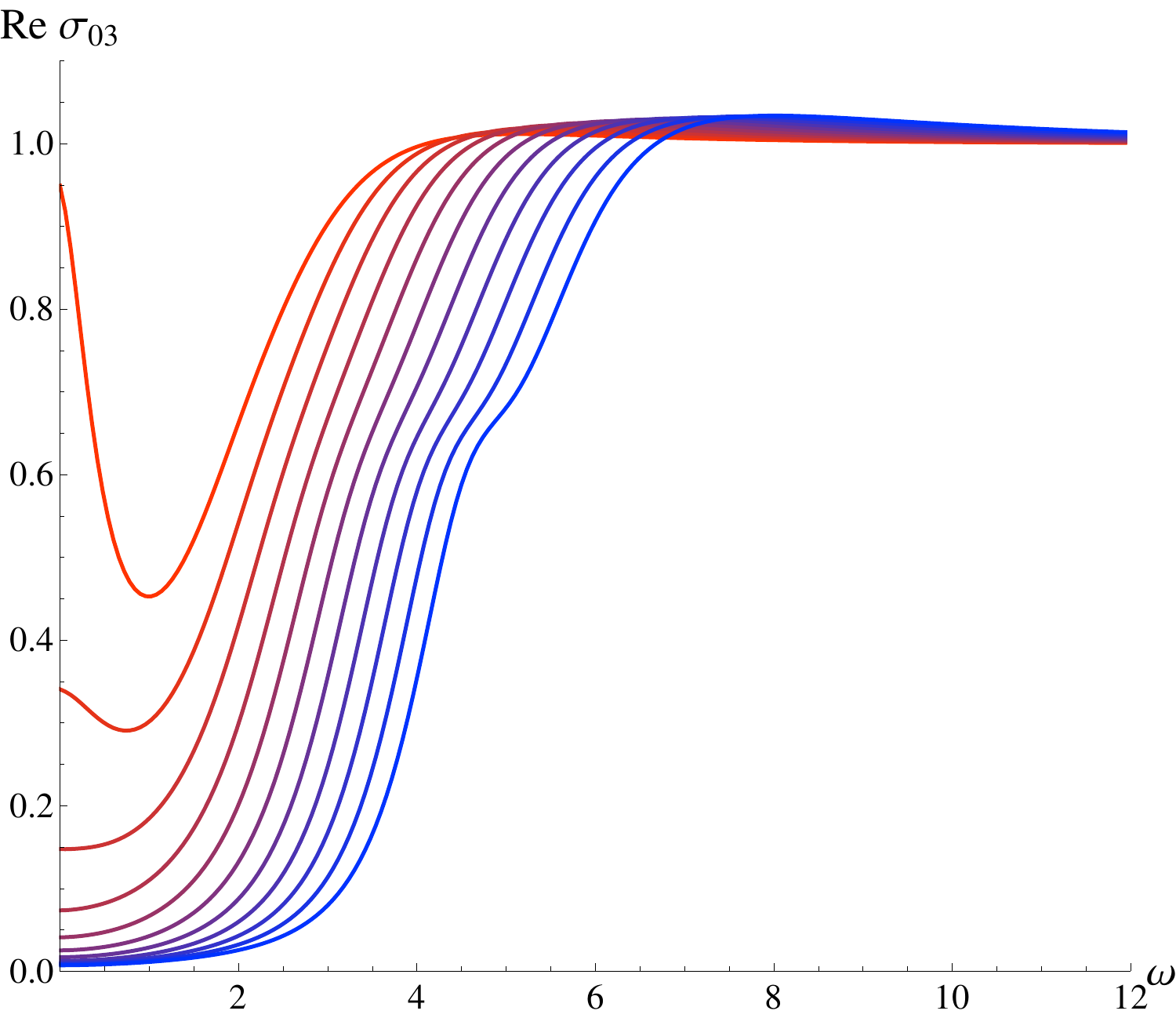}\hfill
\includegraphics[width=225pt]{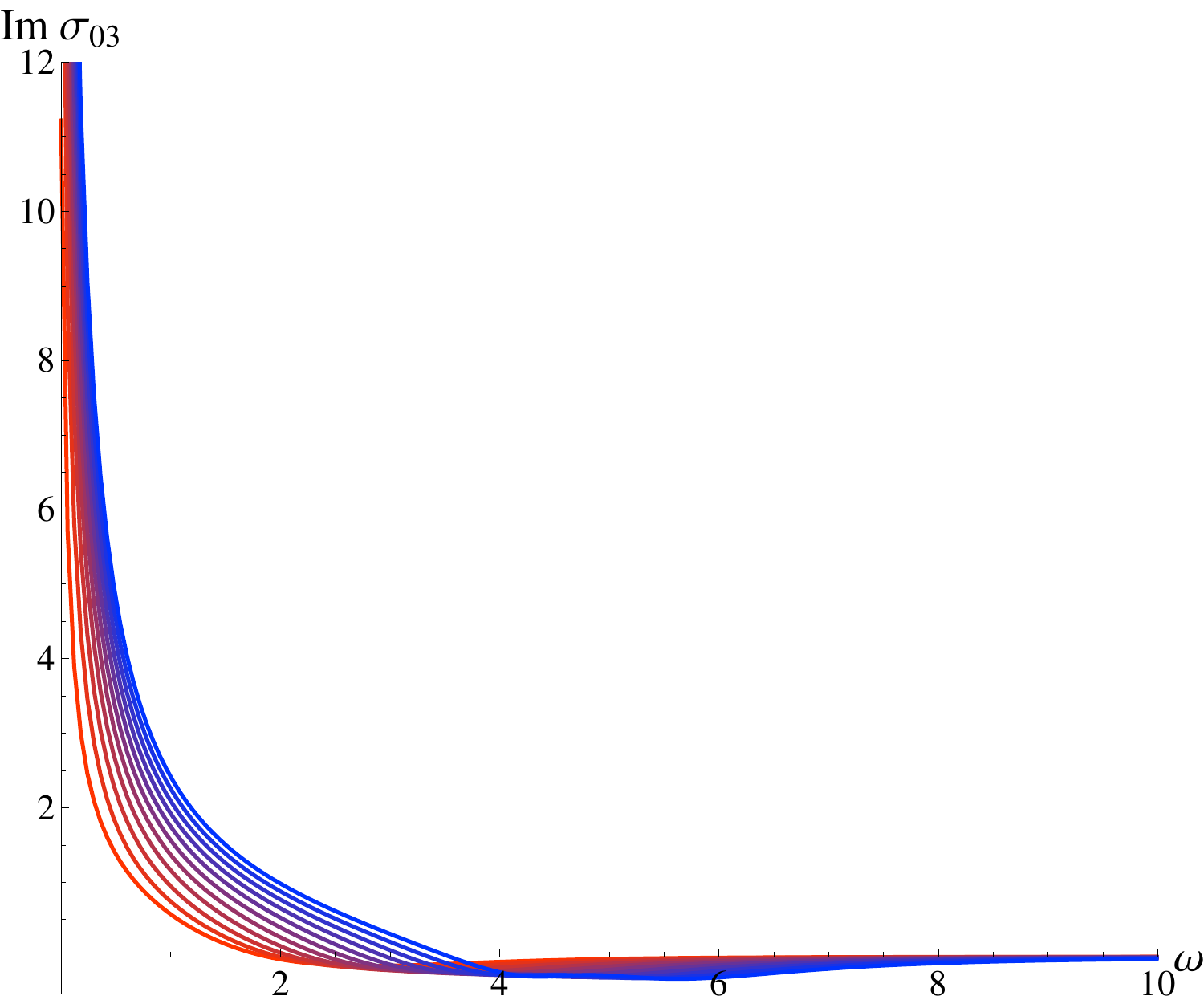}
\caption{Real (left) and imaginary (right) parts of the conductivity for fixed
$S_x/\mu=0.4$. Different lines correspond to different temperatures in the range
$T=0.35 T_c$ (blue) - 0.65$T_c$ (red).}
\label{fig:condhigh}
\end{figure}

Our results show little deviation from what was found at zero superflow. The most
interesting new feature is a low frequency peak which appears due to the
coupling between the gauge and the scalar sectors induced by the superfluid velocity. 
In Figures \ref{fig:condlow} and \ref{fig:condhigh} we present the results for
different values of $S_x/\mu$. As expected the behavior for small
superfluid velocity far from the critical temperature is the same as the one obtained in
\cite{Hartnoll:2008vx}. Close to $T^*$ a bump is generated in the
real part of the conductivity at $\omega \approx 0$. This indicates the existence of a mode
with very small imaginary gap. The mode
responsible for this behavior is the pseudo-diffusive mode described in
\cite{Amado:2009ts}. Due to the conserved $U(1)$ symmetry of the unbroken phase,
there exists a diffusive (gapless) mode in the QNM spectrum of the theory. Once
the symmetry is spontaneously broken, this mode develops a purely imaginary gap
that increases as we lower the temperature. Therefore, for high enough
temperatures below the phase transition, the gap of the pseudo-diffusive mode 
at $k=0$ is very small and
this implies the appearance of a peak at small frequencies in the
conductivity as we can see in the figures. If we lower the temperature, the bump
starts disappearing simply because the gap of the pseudo-diffusive mode
becomes larger. Although this mode was already present in the analysis of the
conductivities without superflow, it is only in our present case that it affects
the conductivity, due to the coupling at zero momentum between the gauge and
scalar sectors mediated by the field $A_x$. The size of the peak is
proportional to the size of that coupling, i.e. it grows with  $S_x/\mu$.

\section{Landau criterion for holographic Type II Goldstone bosons}

In the previous section we studied the lowest lying QNM contained in the
$(0)-(3)$ or $U(1)$ sector of the theory for various values of the superfluid velocity
and arbitrary angle between the momentum and the direction of the superflow. In
this section we extend the analysis to the $(1)-(2)$ sector, which is particular
of the $U(2)$ model of \cite{Amado:2013xya} and contains a type II Goldstone boson in
the spectrum, whose dispersion relation is given by (\ref{eq:reltypeII}) in the
hydrodynamic limit.

The equations describing the system can be found in Appendix B. In this case we choose the momentum to lie always in the direction opposite to the superflow, because as we will see this mode is always unstable. Along with the scalar perturbations prescribed by (\ref{eq:pertsc}) we have to consider the following gauge perturbations in the $(1)-(2)$ sector
\begin{align}
A^{(1)} = a^{(1)}_t(t,r,x)dt +  a^{(1)}_x(t,r,x)dx \,,\nonumber\\
A^{(2)} = a^{(2)}_t(t,r,x)dt +  a^{(2)}_x(t,r,x)dx \,.
\end{align}

Again we use the determinant method of \cite{Kaminski:2009dh} to find the QNMs in this sector. Our
results are summarized in Figure \ref{fig:figure}, where the dispersion relation
for the lowest QNM mode is shown at a particular superfluid velocity. We checked that
the result is qualitatively the same 
for arbitrary $S_x/\mu$. 

The type II Goldstone mode becomes
unstable for arbitrarily small superfluid velocities and temperatures below
$\tilde{T}$. However,
an important difference arises with respect to the $U(1)$ sector. The tachyonic mode does not become stable at any
temperature below $\tilde{T}$, contrary to the situation in the $(0)-(3)$
sector, there is no analogous of $T^*$ in this sector. This behavior can be
easily interpreted as a reflection of the Landau criterion of superfluidity in
our holographic setup: according to (\ref{eq:landaucrit}), the critical velocity
is zero in any system featuring type II Goldstone bosons, hence for any $T<\tilde{T}$
the superfluid phase is not stable at any finite superfluid velocity. In addition notice that the maximum in the imaginary part occurs at higher values of the
momentum as we lower the temperature. In fact as we can see
from the figure, lowering the temperature below $\tilde T$ the maximum in
$\Im(\omega)$ first increases but then starts to decrease again as the temperature
is lowered. At the same time it moves out to ever larger values of the momentum. 

\begin{figure}[htp!] 
\centering
\includegraphics[width=225pt]{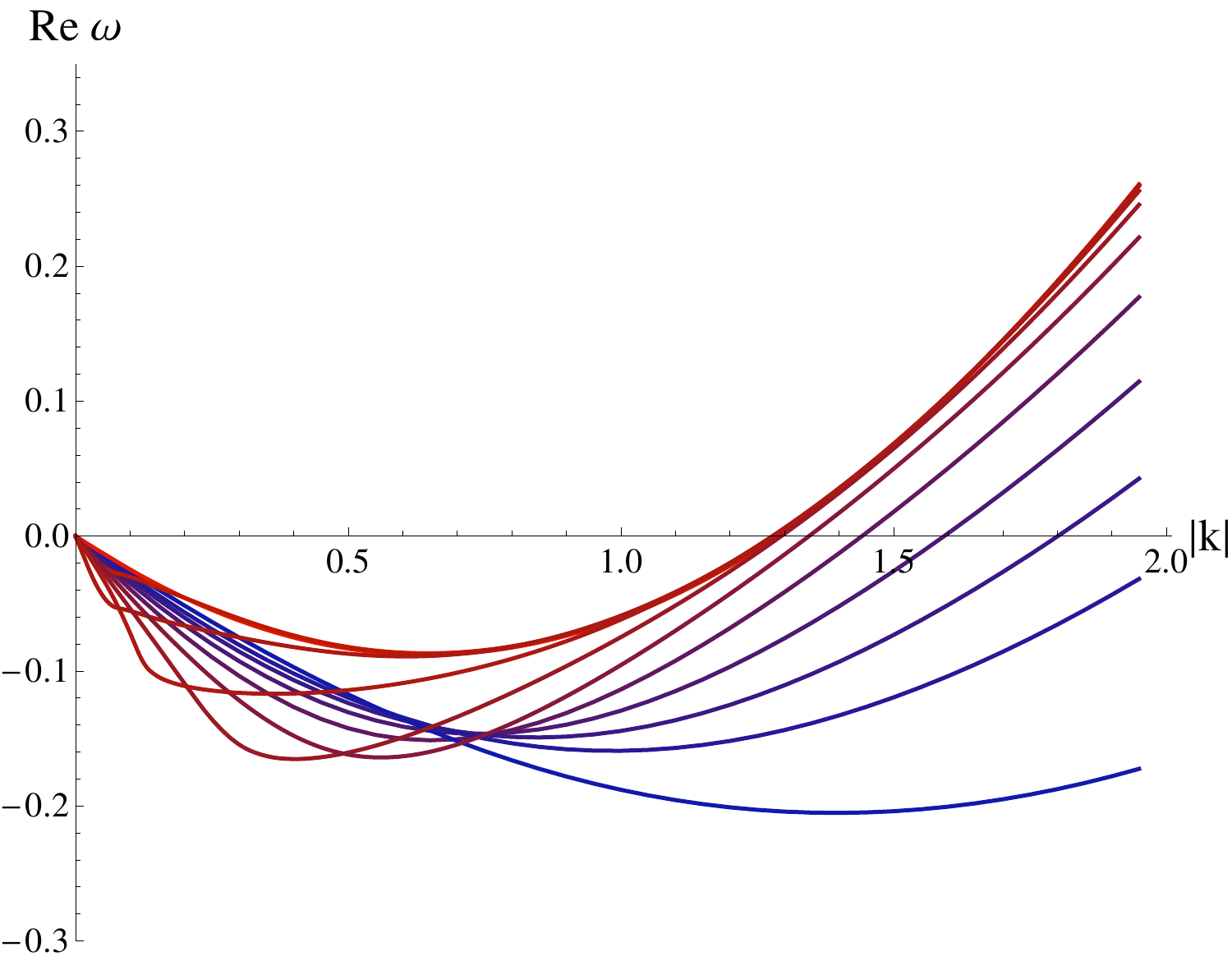}\hfill
\includegraphics[width=225pt]{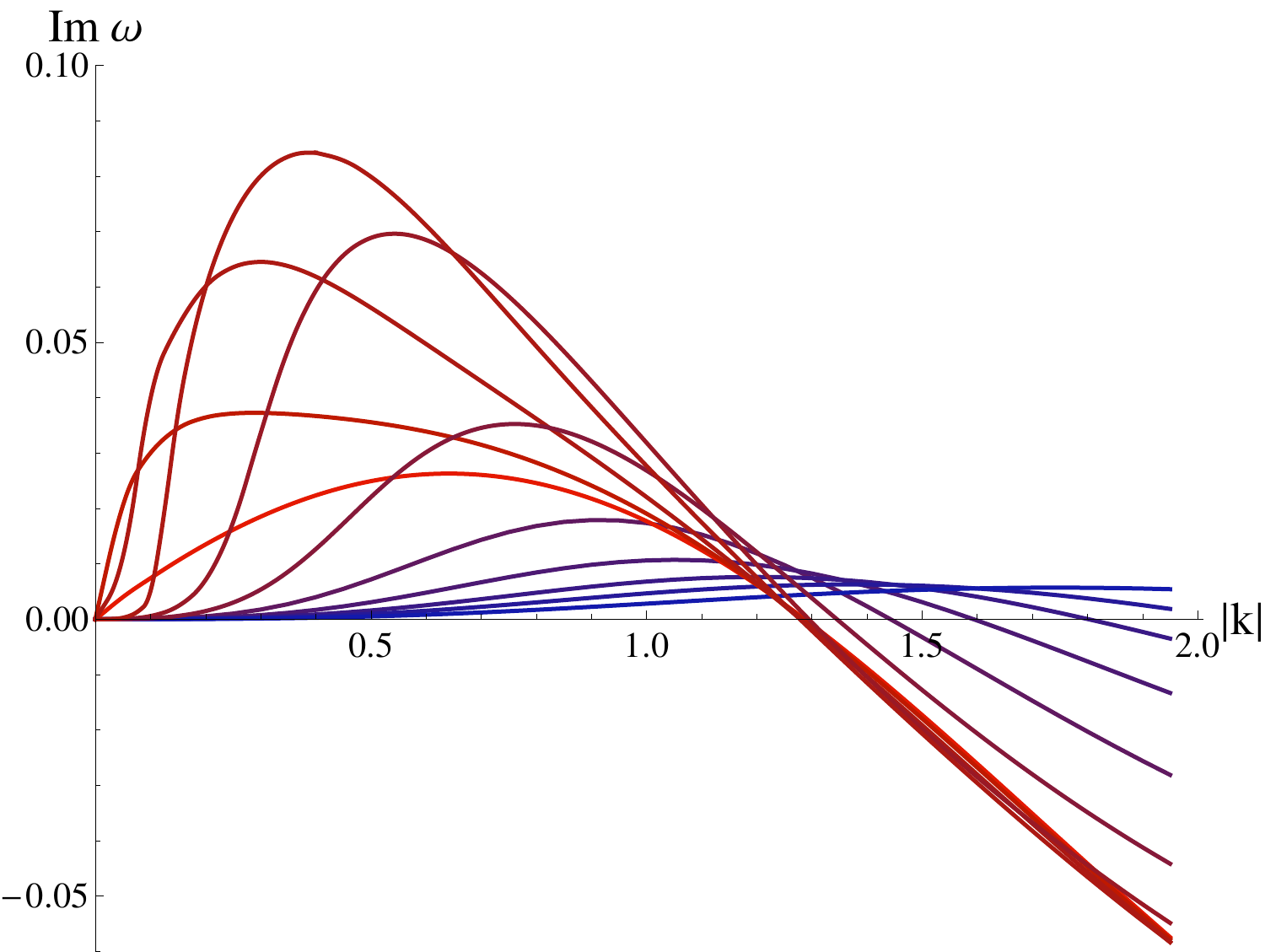}
\caption{Real (left) and imaginary (right) parts of the dispersion relation of
the lowest QNM of the $(1)-(2)$ sector in the gauged model for fixed
$S_x/\mu=0.15$ and a range of temperatures from $T=\tilde{T}=0.95T_c$ (red) to
$T=0.45T_c$ (blue) and momentum anti-parallel to the superfluid velocity.}
\label{fig:figure}
\end{figure}

\begin{figure}[htp!] 
\centering
\includegraphics[width=225pt]{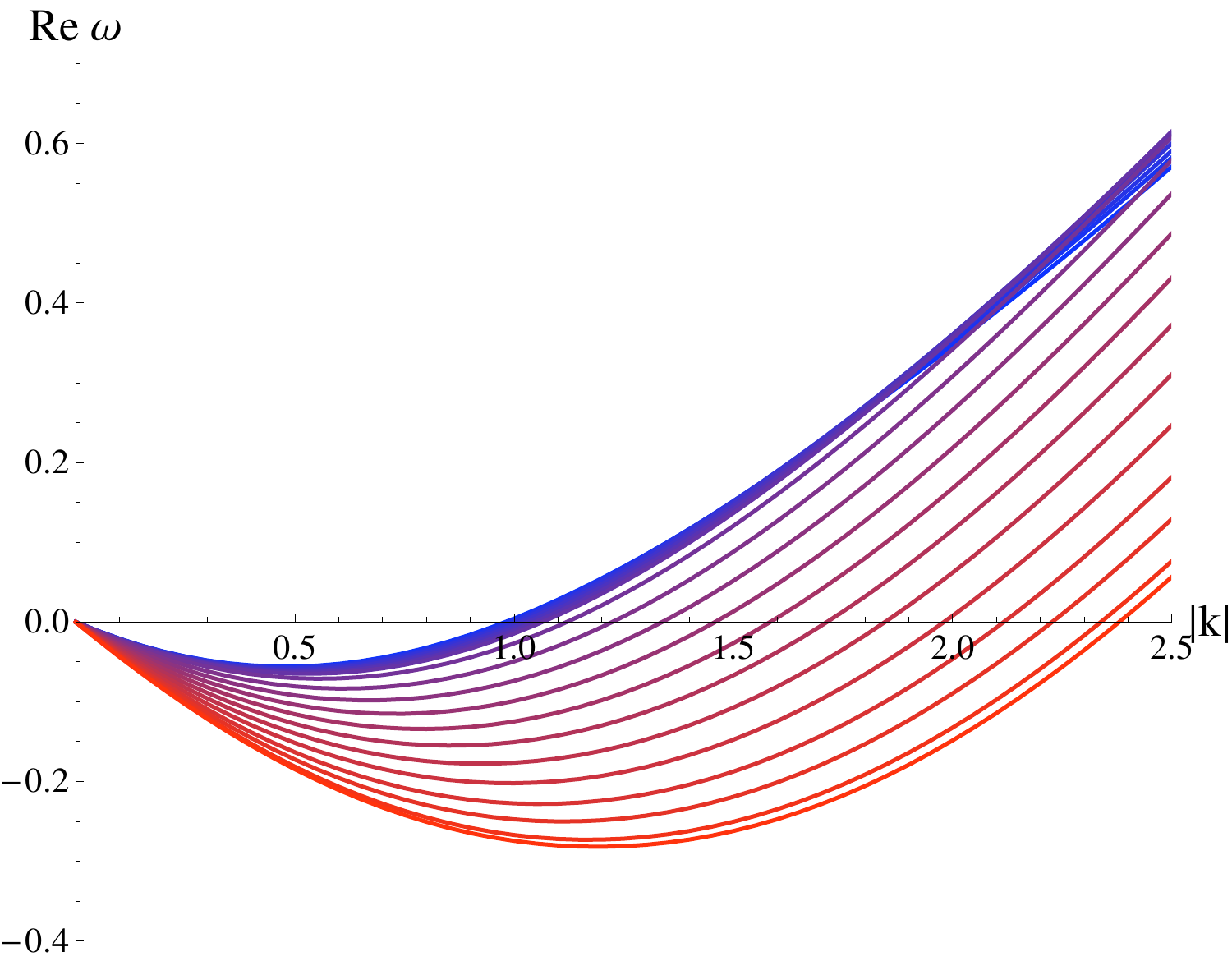}\hfill
\includegraphics[width=225pt]{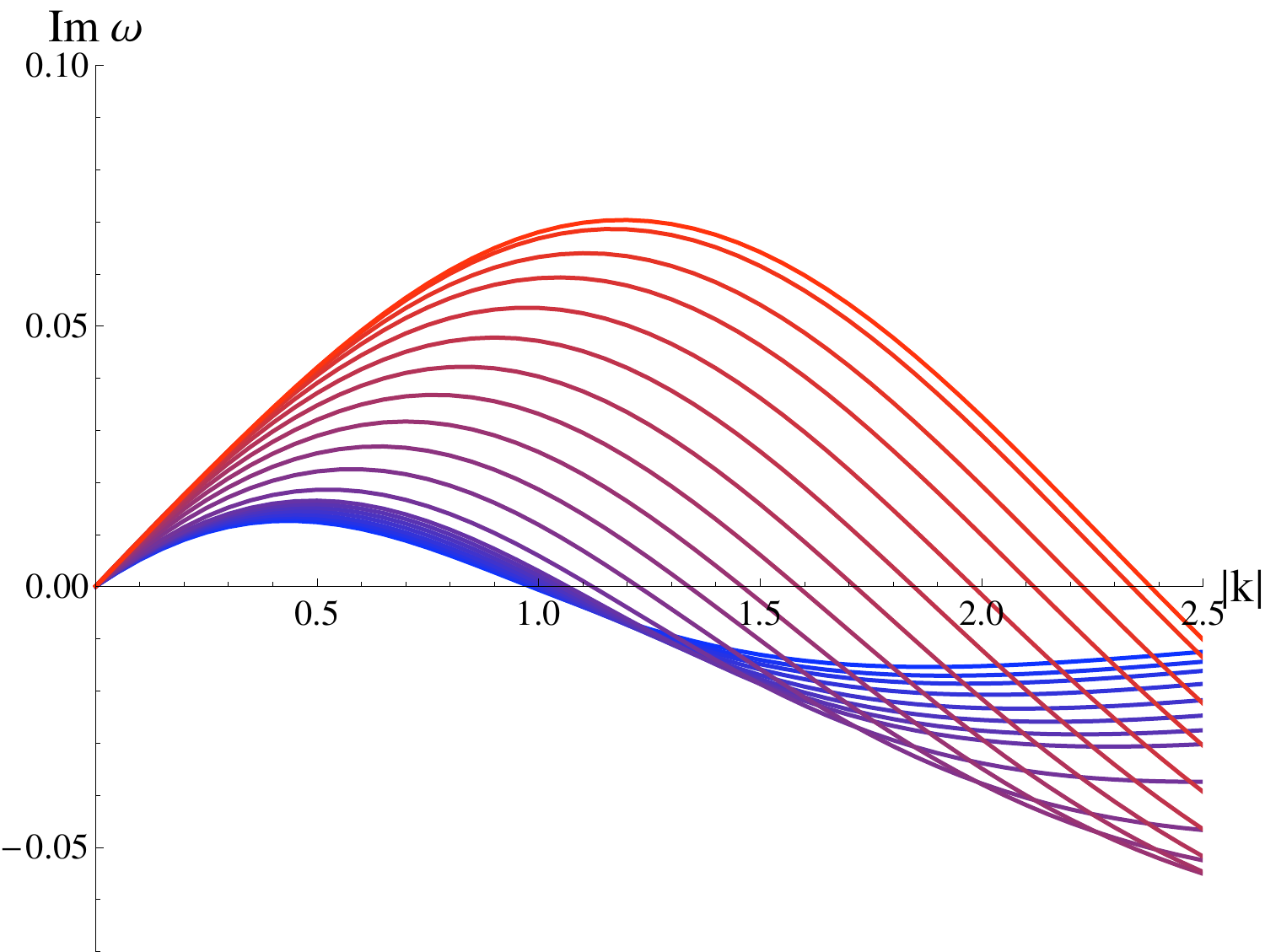}
\caption{Real (left) and imaginary (right) parts of the dispersion relation of
the lowest QNM in the $(1)-(2)$ sector of the ungauged model for fixed
$S_x/\mu=0.25$ and a range of temperatures from $T=\tilde{T}=0.853T_c$ (red) to
$T=0.306 T_c$ (blue). Momentum is taken anti-parallel to the superfluid
velocity.}
\label{fig:figureung}
\end{figure}

Note that plots analogous to Figures \ref{fig:pica0} and \ref{fig:pica1} do not
make any sense in the $U(2)$ model, since the $(1)-(2)$ sector is unstable at any
temperature we have been able to check.

\subsection{Ungauged model}

In \cite{Amado:2013xya} an ungauged model was defined in which there were
no dynamical $SU(2)$ gauge fields in the bulk. This model has a global 
$SU(2)$ symmetry and a local $U(1)$ symmetry. The dual field theory
does therefore not possess the generators of the $SU(2)$ symmetry in its
operator  spectrum. Nevertheless, as shown in \cite{Amado:2013xya} a somewhat
unexpected type II Goldstone mode is present in the QNM spectrum of the model.

The ungauged model is basically given by the same action (\ref{eq:Lagrmodel}) once we keep only the overall $U(1)$ gauge field. Actually it corresponds to the simple $U(1)$ model with two scalar fields with degenerate mass and therefore has an accidental $SU(2)$ global symmetry.

The background solution is again that of the $U(1)$ superfluid, hence the superflow solution can be accommodated also in the ungauged model. The difference
is that the type II Goldstone mode appears now in the fluctuations of the upper component of the scalar field  $\eta$, whose equation of motion reads
\begin{equation}
f \eta''+ \left(f' +\frac{2 f}{r}\right)\eta' + \left( \frac{(\omega + A_0)^2}{f} -\frac{( k- A_x)^2}{r^2} -m^2 \right) \eta=0\,,
\end{equation}
and is completely decoupled of all other field fluctuations. As noticed in \cite{Amado:2013xya} the
change of the background due to the condensate is enough to trigger the appearance of the type II Goldstone.

It is remarkable that in the ungauged model the type II Goldstone mode is still unstable at any temperature below $\tilde{T}$ for any value of the superfluid velocity. Notice that not including conserved currents for the $SU(2)$ symmetry, the model does not satisfy all theorems on existence of Goldstone bosons \cite{Amado:2013xya}. However, the Landau criterion of stability is still valid.

The ungauged model presents a qualitative difference with respect to the gauged model. The value of the momentum at the maximum now decreases as we lower the temperature. This is shown in Figure \ref{fig:figureung}, where the dispersion relation of the type II Goldstone at fixed superfluid velocity and for a long range of temperatures is plotted. For arbitrary values of the superfluid velocity we obtained analogous results.

\section{Conclusions}

We have analyzed the holographic realization of the Landau criterion of
superfluidity. The study was motivated by the appearance of type II Goldstone bosons in
the model (\ref{eq:Lagrmodel}). 
The quadratic nature of the dispersion relation of the type II Goldstone mode
should be responsible for driving the
system out of the superfluid phase for arbitrarily small superfluid velocity.

Taking advantage of the fact that the usual $U(1)$ holographic s-wave
superconductor is contained in (\ref{eq:Lagrmodel}), we have revisited the
Landau criterion for holographic type I Goldstone modes. When addressing the 
question of the stability of the condensate at finite
superfluid velocity the analysis of the free energy does not give the
correct answer. The QNM spectrum contains a tachyonic mode at finite momentum for
temperatures $T^* <T< \tilde{T}$. As defined $\tilde{T}$ is the temperature at which 
free energies of the normal and condensate phase coincide. In contrast, $T^*$ is the temperature where the tachyonic instability arises.
Hence, the homogeneous superfluid is stable only for $T<T^*$, see
Figure \ref{fig:phasespace}. The results for the sound velocity as a function of
the angle $\gamma$ between the propagation direction and the superfluid velocity,
depicted in Figures \ref{fig:pica0} and \ref{fig:pica1}, are perfectly
consistent with this statement: at $T= T^*$ and $\gamma=\pi$ the velocity of
sound vanishes. This condition can be seen to be equivalent to the Landau
criterion and  signals the existence of a critical velocity above which the
superfluid is not stable anymore. 

Since the maximum of the imaginary part of the unstable mode has non-vanishing
wave number it is natural to suggest that there might be another, spatially modulated
phase for $T>T^*$. 
The nature or this inhomogeneous phase is however unknown and we leave its explicit
construction of even the question of its very existence for future research. 

We have also computed the longitudinal conductivities for various
superfluid velocities. As far as we know, they have not been computed before. We see a
peak at $\omega =0$, due to the coupling with the spatial component of the gauge field $A_x$.
The peak decreases as
we lower the temperature until it gets completely suppressed (Figure
\ref{fig:condlow}). We believe that this enhancement of the DC conductivity is
caused by the gap of the pseudo-diffusive mode \cite{Amado:2009ts, Amado:2013xya} 
which in the presence of superfluid velocity is formed due to the coupling between the gauge 
and scalar sectors that takes place even at $k=0$. 

Moving to the $(1)-(2)$ sector, we  worked out the impact of the superflow
on  the type II Goldstone mode. We found that the Landau criterion is effective
for arbitrarily small superfluid velocity as depicted in Figure \ref{fig:figure}. 
The tachyon persists for the whole range
of temperatures and (finite) superfluid velocities we have been able to analyze.
Hence, we conclude that the critical superfluid velocity for this sector vanishes, in
complete accordance with the Landau criterion applied to modes with dispersion
relation $\omega \propto k^2$. An analogous result holds for the type II Goldstone
mode in the ungauged model.      

\section*{Acknowledgements}
 We have profited a lot from discussions with V. Giraldo and A. Schmitt. A. J.
would like to thank Ioannis Papadimitriou for useful discussions.  L.M. wants to thank the Imperial College London for their
hospitality during his research visit, specially J. P. Gauntlett, C. Pantelidou
and G. De Nadai Sowrey.
This work has been supported by MEC and FEDER grant FPA2012-32828,
Consolider Ingenio Programme CPAN (CSD2007-00042), Comunidad de Madrid HEP-HACOS
S2009/ESP-1473
and MINECO Centro de excelencia Severo Ochoa Program under grant SEV-2012-0249.
I. A. is supported by the Israel Science Foundation under grants no. 392/09 and
495/11. L.M. has been supported by FPI-fellowship
BES-2010-041571. A. J. is supported by FPU fellowship AP2010-5686. 
D. A. thanks the FRont Of pro-Galician Scientists for unconditional support.

\newpage
\begin{appendix}
\section{Fluctuation equations in the $(0)-(3)$ sector}

The fluctuations in the $U(1)$ theory or the $(0)-(3)$ sector contain the zeroth
and third color sectors of the gauge field and the lower
component of the scalar field $\sigma=\rho+i\delta$. The equations of motion for
an arbitrary direction of the momentum then read
\begin{align}
\nonumber 0=&f\rho'' +\left(f' +\frac{2f}{r}\right) \rho'
+\left(\frac{\omega^2}{f}+\frac{A_0^2}{f}- \frac{A^2_x}{r^2}-
\frac{|k|^2}{r^2}  -m^2\right)\rho + \frac{2 i\omega A_0}{f}\delta +2
a^{(-)}_t\Psi\frac{A_0}{f}\\
&- 2\frac{a^{(-)}_x}{r^2}\Psi A_x +|k|\cos(\gamma)\frac{2i}{r^2}A_x\delta\,,\\
\nonumber 0=&f\delta'' +\left(f' +\frac{2f}{r}\right) \delta'
+\left(\frac{\omega^2}{f}+\frac{A_0^2}{f}- \frac{A^2_x}{r^2} -
\frac{|k|^2}{r^2}  -m^2\right)\delta - \frac{2 i\omega A_0}{f}\rho - i
\Psi\omega\frac{a^{(-)}_t}{f} \\
&- |k|\cos(\gamma)\frac{2i}{r^2}A_x\rho -
|k|\cos(\gamma) \frac{i}{r^2} \Psi a^{(-)}_x- |k|\sin(\gamma) \frac{i}{r^2} \Psi
a^{(-)}_y\,,\\
\nonumber 0=&f a''^{(-)}_t + \frac{2f}{r}
a'^{(-)}_t-\left(\frac{|k|^2}{r^2}+2\Psi^2\right) a^{(-)}_t - \frac{\omega
|k|}{r^2} \cos(\gamma) a^{(-)}_x- \frac{\omega |k|}{r^2} \sin(\gamma)
a^{(-)}_y\\
&-4\Psi A_0 \rho - 2i\omega\Psi \delta\,,\\
\label{QNM03x} \nonumber 0=& f a''^{(-)}_x +f' a'^{(-)}_x +\left(\frac{\omega^2
}{f} -2 \Psi^2\right)a^{(-)}_x +\frac{\omega |k|}{f}\cos(\gamma)
a^{(-)}_t+2i|k|\cos(\gamma) \Psi \delta  \\
&-4 \Psi \rho A_x- \frac{|k|^2\sin^2(\gamma)}{r^2}a^{(-)}_x+
\frac{|k|^2\cos(\gamma)\sin(\gamma)}{r^2}a^{(-)}_y\,,\\
\label{QNM03y} \nonumber 0=& f a''^{(-)}_y +f' a'^{(-)}_y +\left(\frac{\omega^2
}{f} -2 \Psi^2\right)a^{(-)}_y +\frac{\omega |k|}{f}\sin(\gamma)
a^{(-)}_t+2i|k|\sin(\gamma) \Psi \delta \\
&- \frac{|k|^2 \cos^2(\gamma)}{r^2}a^{(-)}_y+ \frac{|k|^2\cos(\gamma)
\sin(\gamma)}{r^2}a^{(-)}_x\,,
\end{align}
and the constraint
\begin{align}
0=\frac{i \omega}{f}a'^{(-)}_t +\frac{i|k|}{r^2} \cos(\gamma) a'^{(-)}_x
+\frac{i|k|}{r^2} \sin(\gamma) a'^{(-)}_y+ 2 \Psi' \delta - 2\Psi \delta'\,,
\end{align}
where we have used $k_x = |k| \cos(\gamma)$, $k_y = |k| \sin(\gamma)$ . The general pure gauge
solution in this sector is
\begin{align}
\delta = i \lambda \Psi; \hspace{0.5cm} \rho = 0; \hspace{0.5cm} a^{(-)}_t =
\lambda \omega; \hspace{0.5cm} a^{(-)}_x=- \lambda |k| \cos(\gamma);
\hspace{0.5cm} a^{(-)}_y=- \lambda |k| \sin(\gamma)\,,
\end{align} 
where $\lambda$ is an arbitrary constant.

\section{Fluctuation equations in the $(1)-(2)$ sector}

The perturbations in the $(1)-(2)$ sector of the $U(2)$ theory include the
fluctuations of the upper component of the scalar field, $\eta=\alpha+i\beta$,
along with that sector of the gauge field. For momentum in the opposite direction of the superflow, the equations of motion read
\begin{align}
\nonumber \label{eqna1xk}0&=fa''^{(1)}_x +f'a'^{(1)}_x+
\left(\frac{\omega^2}{f}- \Psi^2 + \frac{\left(A_t^{(3)}\right)^2}{f}\right)a^{(1)}_x -
2i\frac{A_t^{(3)} \omega}{f}a^{(2)}_x + i \omega \frac{A^{(3)}_x}{f} a^{(2)}_t\\
&-\frac{A_t^{(3)} A^{(3)}_x}{f}a^{(1)}_t -2 A^{(0)}_x\Psi \alpha + 2 i k \Psi\beta
-\frac{i k A_t^{(3)}}{f}a^{(2)}_t + \frac{\omega k}{f}a^{(1)}_t\,,\\
\nonumber \label{eqna2xk}0&=fa''^{(2)}_x +f'a'^{(2)}_x+ \left(\frac{\omega^2}{f}
-\Psi^2 + \frac{\left(A_t^{(3)}\right)^2}{f}\right)a^{(2)}_x + 2i\frac{A_t^{(3)} \omega
}{f}a^{(1)}_x- i \omega \frac{A^{(3)}_x}{f}
a^{(1)}_t \\ 
&-\frac{A_t^{(3)} A^{(3)}_x}{f}a^{(2)}_t + 2 \Psi A^{(0)}_x \beta +2i k\Psi \alpha+  \frac{i k A_t^{(3)}}{f}a^{(1)}_t 
+\frac{\omega k}{f}a^{(2)}_t\,,\\
\nonumber 0&=fa''^{(1)}_t + \frac{2f}{r} a'^{(1)}_t
-\left(\frac{\left(A_x^{(3)}\right)^2}{r^2}+ \Psi^2+ \frac{k^2}{r^2}\right)a^{(1)}_t
+\frac{A_t^{(3)} A^{(3)}_x}{r^2}a^{(1)}_x - i\omega \frac{A^{(3)}_x}{r^2} a^{(2)}_x
- 2i\omega \beta \Psi \\
&- 2 \phi \Psi \alpha+ \frac{i k A_t^{(3)}}{r^2}a^{(2)}_x- \frac{2i k A^{(3)}_x}{r^2}a^{(2)}_t- \frac{\omega k}{r^2}a^{(1)}_x\,,\\
 \nonumber 0&=fa''^{(2)}_t + \frac{2f}{r} a'^{(2)}_t -
\left(\frac{\left(A_x^{(3)}\right)^2}{r^2}+ \Psi^2+ \frac{k^2}{r^2}\right)a^{(2)}_t
+\frac{A_t^{(3)} A^{(3)}_x}{r^2}a^{(2)}_x + i\omega \frac{A^{(3)}_x}{r^2} a^{(1)}_x
\\
&-\frac{i k A_t^{(3)}}{r^2}a^{(1)}_x+ \frac{2i k A^{(3)}_x}{r^2}a^{(1)}_t- \frac{\omega k}{r^2}a^{(2)}_x 
- 2i\omega \alpha \Psi+ 2 A_t^{(0)} \Psi \beta\,, \\
\nonumber 0&= f\alpha'' + \left(f' +\frac{2f}{r} \right)\alpha' +
\left(\frac{\omega^2}{f} + \frac{\left(A_t^{(0)}+ A_t^{(3)}\right)^2}{4f} - \frac{\left(
A^{(0)}_x+ A^{(3)}_x\right)^2}{4r^2}-\frac{k^2}{r^2}-m^2 \right) \alpha  \\
&+\left(i \omega\left(\frac{A_t^{(0)}
+ A_t^{(3)}}{f}\right) +
\frac{i k}{r^2}\left(A^{(0)}_x+A^{(3)}_x\right)\right) \beta  +
\frac{A_t^{(0)} \Psi}{2 f} a^{(1)}_t - i\omega \frac{\Psi}{2f} a^{(2)}_t \nonumber\\
&-\frac{A^{(0)}_x \Psi}{2r^2} a^{(1)}_x- \frac{i k \Psi}{2r^2}a^{(2)}_x\,,\\
 0&= f\beta'' + \left(f' +\frac{2f}{r} \right)\beta' +
\left(\frac{\omega^2}{f} + \frac{\left(A_t^{(0)}+ A_t^{(3)}\right)^2}{4f} - \frac{\left(
A^{(0)}_x+ A^{(3)}_x\right)^2}{4r^2}-\frac{k^2}{r^2}-m^2 \right) \beta\nonumber\\
&-\left( i \omega\left(\frac{A_t^{(0)} +
A_t^{(3)}}{f}\right) + \frac{i k}{r^2}\left(A^{(0)}_x+
A^{(3)}_x\right)\right) \alpha  - \frac{A_t^{(0)} \Psi}{2 f} a^{(2)}_t - i\omega
\frac{\Psi}{2f} a^{(1)}_t \nonumber\\
&+ \frac{A^{(0)}_x \Psi}{2r^2} a^{(2)}_x- \frac{i k \Psi}{2r^2}a^{(1)}_x\,,
\end{align}
subject to the constraints
\begin{align}
 0= 2 f \left(\Psi \beta' - \Psi' \beta
\right) + a^{(2)}_t A'^{(3)}_t- a'^{(2)}_t A_t^{(3)}+  \frac{f}{r^2} \left(A^{(3)}_x
a'^{(2)}_x - a^{(2)}_x A'^{(3)}_x\right)-i\omega a'^{(1)}_t -
\frac{i k f}{r^2}a'^{(1)}_x \,,\\
 0= 2 f \left(\Psi \alpha' - \Psi' \alpha\right) +  a'^{(1)}_t A_t^{(3)}- a^{(1)}_t A'^{(3)}_t+  \frac{f}{r^2} \left(a^{(1)}_x
A'^{(3)}_x-A^{(3)}_x a'^{(1)}_x\right)-i\omega a'^{(2)}_t -\frac{i k f}{r^2}a'^{(2)}_x\,,
\end{align}
There are two pure gauge solutions in this sector,
\begin{align}
\alpha=0 \,, \quad \beta=i \lambda_1 \Psi/2 \,,\quad a_t^{(1)}=\lambda_1
\omega \,,\quad  a_t^{(2)}=i \lambda_1 A_t^{(3)} \,,\quad a_x^{(1)}=-\lambda_1 k \,,\quad a_x^{(2)}=i \lambda_1 A_x^{(3)}
\,,\\
 \alpha=i \lambda_2 \Psi/2 \,, \quad \beta=0 \,, \quad a_t^{(1)}=-i \lambda_2
A_t^{(3)} ,\quad a_t^{(2)}=\lambda_2 \omega \,,\quad a_x^{(1)}=-i \lambda_2 A_x^{(3)}, \quad  a_x^{(2)}=-\lambda_2 k
\,,
\end{align}
where $\lambda_1$ and $\lambda_2$ are arbitrary constants.

\end{appendix}
 
\bibliography{Supflowv2}{}
\bibliographystyle{utcaps}

\end{document}